\documentclass[aip,jcp,unsortedaddress]{revtex4-1}
\usepackage{amssymb}
\usepackage{amsmath}
\usepackage{graphicx}
\usepackage[sort&compress]{natbib}
\usepackage[dvips]{color}

\begin{document}

\title{Aggregation of amphiphilic polymers in the presence of adhesive small
colloidal particles}
\author{Vladimir A. Baulin}
\affiliation{ICREA, 23 Passeig Lluis Companys, 08010 Barcelona, Spain}
\affiliation{Departament d'Enginyeria Qu\'{\i}mica, Universitat Rovira i Virgili 26 Av.
dels Paisos Catalans, 43007 Tarragona Spain}
\author{Albert Johner}
\affiliation{Institut Charles Sadron 23, rue du Loess 67034 Strasbourg France}
\author{Josep Bonet Avalos}
\affiliation{Departament d'Enginyeria Qu\'{\i}mica, Universitat Rovira i Virgili 26 Av.
dels Paisos Catalans, 43007 Tarragona Spain}
\date{\today }

\begin{abstract}
The interaction of amphiphilic polymers with small colloids, capable to
reversibly stick onto the chains, is studied. Adhesive small colloids in
solution are able to dynamically bind two polymer segments. This association
leads to topological changes in the polymer network configurations, such as
looping and cross-linking, although the reversible adhesion permits the
colloid to slide along the chain backbone. Previous analyses only consider
static topologies in the chain network. We show that the sliding degree of
freedom ensures the dominance of small loops, over other structures, giving
rise to a new perspective in the analysis of the problem. The results are
applied to the analysis of the equilibrium between colloidal particles and
star polymers, as well as to block copolymer micelles. The results are
relevant for the reversible adsorption of silica particles onto hydrophilic
polymers, used in the process of formation of mesoporous materials of the
type SBA or MCM, cross-linked cyclodextrin molecules threading on the
polymers and forming the structures known as polyrotaxanes. Adhesion of
colloids on the corona of the latter induce micellization and growth of
larger micelles as the number of colloids increases, in agreement with
experimental data.
\end{abstract}

\maketitle

\section{Introduction}

The interaction between polymers and small solute molecules, such as
surfactants or colloidal particles, attracts a great interest due to wide
industrial applications and their biological significance \cite{Hamley,Reiss}%
. For example, polymers are used to control the stability of colloidal
suspensions \cite{Halperin,Napper}, block copolymer micelles are employed
for targeted delivery of small colloids \cite{Kataoka}, but essentially the
interaction of biopolymers with proteins or membrane phospholipids is of
fundamental importance \cite{Alberts}. In addition, the interactions of
hydrophilic polymers and brushes with \textit{large} colloidal particles
\cite{Diez,Gelbart,Lykos} or micelles \cite{Diamant} has been a subject of
intensive research.

The presence of hydrophilic polymers induces the formation of
polymer--surfactant aggregates in the form of micelles covered by polymer
chains \cite{Diamant}. This cooperative association usually happens below
the CMC of the surfactants. Large colloidal particles form aggregates with
polymers where hydrophilic polymer chains are wrapped around colloidal
particles \cite{Diez,Gelbart}. The polymer chains can link different
colloids between each other, thus resulting in the clustering of colloids
\cite{Lykos} or gel formation \cite{Diez}. In turn, several polymer chains
can adsorb on the same colloidal particle forming the hydrophilic corona and
stabilizing it. A notable feature of such aggregates is that if adsorption
is reversible, colloidal particles can effectively slide along the chain. In
the case of many long chains adsorbed on the same colloidal particle, this
sliding degree of freedom results in the formation of star-shaped aggregates
with annealed number of arms \cite{Diez}.

However, the interaction of polymers with \textit{small} colloidal particles
has received relatively little attention. In this paper we precisely focus
our interest on interactions of hydrophilic polymers with \textit{small}
colloidal particles that can reversibly stick to hydrophilic parts of the
polymer backbone. In contrast to large colloids, the polymers do not adsorb
onto the particles but only adhere onto a few active sites, i.e., the small
colloids induce topological changes in the polymer configurations, promoting
looping, cross-linking, and interconnections between the chains. In this
context, the following situations, which are relevant in practice, can be
mentioned: (i) The driving force for the formation of mesoporous
silica-based materials type SBA or MCM \cite{dynmesop} is the strong
interaction of hydrophilic polymers with silica particles, e.g. TEOS, and
the subsequent formation of self-assembled structures. The possibility of
tuning and controlling the design of the resulting structures is a
challenging task. Silicon alkoxides in water form Si-OH compounds, but also
different pairs or oligomers of silicon, namely Si-O-Si \cite{Iler}. Silica
particles form hydrogen bonds with hydrophilic polymer's backbone \cite%
{SilPEO} . Since the hydrogen bonds are relatively weak, the silica
particles can effectively slide along the chain by breaking and re-forming
new hydrogen bonds. Thus, we can consider a silica particle as being either
a sliding link between different chains or a sliding loop on the same chain.
Addition of silica particles to the solution of block copolymer micelles
leads to the adsorption of silica particles on the coronas of micelles \cite%
{dynmesop}. The practical example of such systems is an aqueous solution of
TEOS with common triblock copolymers Pluronic P123 (EO$_{20}$-PO$_{70}$-EO$%
_{20}$) or Pluronic F127 (EO$_{106}$-PO$_{70}$-EO$_{106}$). The micelles and
self-assembled ordered structures of these polymers serve as precursors for
the pores in the mesoporous materials. The diameter of hydrophobic cores of
micelles determines the size of the pores, while the structure of the
polymeric corona is responsible for the microporosity of the material \cite%
{Micropor}. Therefore, the control of the adhesion of the silica particles
into the coronas can lead to the control of the core of the micelle and,
eventually to the control of the properties of the final porous material.
(ii) A cone shaped molecules with a hydrophobic cavity known as
cyclodextrins \cite{HaradaAcc,HaradaCoo,HaradaLi3} can spontaneously thread
on linear hydrophilic polymers. Polymer chains dressed with necklaces of
cyclodextrins bear the special name of polyrotaxanes. Since two molecules of
cyclodextrins can be covalently bound together, the resulting molecule will
represent two linked rings. Such molecules can either link two chains or
form a loop on the same chain, providing a sliding degree of freedom for the
chain to move inside the rings. In previous works the effect of these
sliding links for the polymers grafted to a surface \cite{BaulinLayers} and
the micelles with sliding coronas \cite{BaulinMic} has been studied.

\begin{figure}[ht]
\begin{center}
\includegraphics[width=8.5cm]{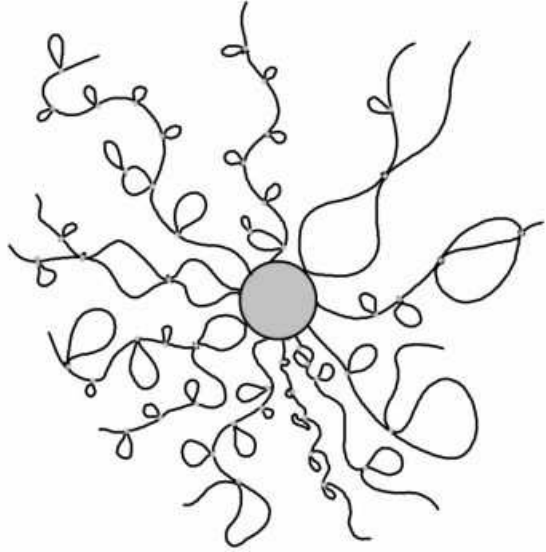}
\end{center}
\caption{Schematic representation of a block copolymer micelle with small
colloids in the corona. This sketch shows possible topologies addressed in
this work.}
\label{Fig:scketch}
\end{figure}

In this paper we introduce a formalism suitable for a qualitative
description of these mentioned processes, in which flexible polymers
interact with small colloidal particles or other agents, that induce binding
with the polymers. A complete statistical mechanical description of these
processes is of extreme complexity due to the changing topology of the
system due to the adhesion of the colloids in different chain backbones and,
therefore, we will here focus on the qualitative trends that a scaling
theory can reveal. Formally, the complex topology of the polymeric network,
produced by the presence of the binding agents as shown in Figure \ref%
{Fig:scketch}, permits us to use known results of polymeric networks of
fixed topology, for which the relevant scaling exponents are known \cite%
{Duplantier}. The challenging objective of our work is thus to derive the
partition function for a system where the \textit{vertices} of the polymeric
network can slide along chain backbones and, therefore, the topology of the
system is not fixed. This is precisely the case of the physical situations
that we have described so far, and that cannot be trivially addressed by a
simple mapping of the fixed topology results, as done in other context \cite%
{Mukamel,Metzler1,Metzler3}. An important consequence of this fact is that,
in the integration of this degree of freedom, the system passes through
regions with different topology, characterized by different sets of scaling
exponents. In the next section we present what, to the best of our
knowledge, is the first derivation of this effect. Therefore, in the
following we formulate our scaling model and address the question of the
calculation of the singular non-trivial contribution to the entropy of a
system of hydrophilic polymers with small colloidal particles forming
different mobile polymer architectures. With the construction of this
singular part of the partition functions, according to the scaling theory,
we find the most favorable conformations of the system and, as a
consequence, the distribution of colloids along the chains, for instance.

Within this formalism we can calculate the regimes of loading of
silica particles in linear polymers, star polymers, and in
micelles of block copolymers, and permits us to give a theoretical
justification of several experimentally observed facts. We believe
that our approach can be useful in biophysical applications in
systems like DNA chains interacting with proteins related to gene
regulation models, where equilibrium analyses have been carried
out\cite{Vilar}, and also the effect of looping on the transport
of proteins along the chain has been analyzed \cite{Berg,Broek}.
Related analyses can be carried out to study the problem of DNA
denaturation \cite{Poland} and, effectively, have been done to
study the equilibrium shapes of knots in
chains\cite{Metzler5,Ercolini}.

The paper is organized as follows. In section \ref{secmodel}, we introduce
the model of the system and perform the scaling analysis. Section \ref{seceq}
is devoted to the study of the equilibrium between polymeric systems and
colloidal particles, in particular, colloids with star polymers and colloids
with block copolymer micelles. Finally, section \ref{secconclusion} is
devoted to the conclusions that we have drawn from this work.

\section{Model \label{secmodel}}

In this section we propose the scaling form of the partition function for
systems, polymeric networks, of different topology. Such a partition
function contains a trivial energetic contribution arisen from the
short-range interaction between the binding agents and polymer segments.
However, the entropic contribution strongly depends on the topology of the
polymeric network resulting from that interaction, together with the
topological effects that are related to the topology of the polymer itself,
as in the case of star polymers. To that purpose, we refer to the results
obtained by Duplantier \cite{Duplantier1,Duplantier}, for the partition
function of polymer networks of arbitrary topology. Such a partition
function has a superscaling form $Z\sim s^{N}N^{\gamma _{\mathcal{L}%
}-1}g(n_{1}/N,n_{2}/N,...,n_{k}/N)$, where $N$ is the total length (total
number of monomers) of a network, $s$ is a non-universal geometrical
constant, $n_{1},...,n_{k}$ are the lengths of different polymeric threads
between two crosslinks, or between a crosslink and a dangling end, such that
$\sum n_{j}=N$, and $g$ is an unknown function. $\gamma _{\mathcal{L}}$ is
the universal exponent which does not depend on the type of interactions and
is determined only by the topology of the system. It is given by \cite%
{Duplantier}
\begin{equation}
\gamma _{\mathcal{L}}-1=-d\nu q_{loop}+\sum_{k\geq 1}p_{k}\sigma _{k}
\label{genexpr}
\end{equation}%
where $q_{loop}$ is the number of independent loops, $p_{k}$ is the number
of vertices with $k$ legs, $d$ is the dimension of the space, $\nu $ is the
Flory exponent, associated to the radius of gyration of a self-avoiding
walk, $\sigma _{k}$ is the exponent corresponding to a vertice with $k$
-legs. The universal exponents $\nu $ and $\sigma _{k}$ are known exactly
for $d=2$ and $d\geq 4$ and numerically for $d=3$. In the following we use
the values of $\sigma _{k}$ obtained from the simulation results for
critical exponents of star polymers \cite{Grassberger}. Although a general
form of the function $g(n_{1}/N,n_{2}/N,...,n_{k}/N)$ is not analytically
known, the scaling behavior of the system can be derived from the scaling
form of $g$ when some of its arguments go either to $0$ or to $1$. This
scaling form gives the correct asymptotes in these limiting cases. Due to
the fact that our interest lies only in the derivation of the scaling
behavior of the overall system, we will avoid the explicit construction of
the crossover functions between different regimes, characterized by
different topologies, by basically assuming that the crossover function is
of the order of a constant and matching these constants at the crossover
region between different topologies, as shown shortly. This method allows us
to derive an approximate description over the whole range of topological
configurations of the system. From a physical point of view $N$ as well as
series of $n_1, n_2, \dots$, representing the size of the branches of the
polymer network represent the number of Kuhn segments of each polymeric
thread. Furthermore, the validity of the scaling analysis is restricted to a
system with $N$ being very large. In particular, Monte Carlo simulations on
isolated polymers \cite{Binder,Everaersts} indicate that the scaling limit
is reached for sizes exceeding hundreds of Kuhn segments, and therefore our
analysis will be restricted to polymeric systems of at least this size. With
regard to the colloidal particle, we will consider that its size is slightly
larger than the Kuhn segment but much smaller than the overall size of the
polymeric network.

We base our model on the following assumptions regarding the polymer--small
colloids interactions: (i) one sliding sticker can bound several polymer
units either on the same chain or different chains; (ii) the links are
reversible and, hence, the reversibility allows for an effective sliding of
the sticker along the chain; (iii) we assume steric repulsion between small
colloids. Then the colloids and polymers bind together they define a network
whose topology can change due to the sliding degree of freedom of the
colloid. It is important to realize, however, that the \textit{sliding} can
physically take place by reversibly breaking and reforming bonds between the
polymer and the colloid, regardless whether the colloid completely unbinds
from the chain or it slides from site to site, since both are dynamic
processes whose effect is accounted for by the equilibrium statistical
mechanical treatment. Hence the challenging objective of this work is to
determine the partition function of a system with fixed number of polymers
and binding agents, taking into account the degree of freedom of the motion
of the colloid along the chain backbone and, at the same time, correctly
accounting for the topological changes in the network originated by this
degree of freedom. Up to the best of our knowledge, this task has not been
carried out up to date.

In what follows we will apply the formalism to systems of increasing
complexity, emphasizing the details of the calculations in the simpler
systems and making the extension to more complex systems.

\subsection{Stickers in linear chains}

\begin{figure}[ht]
\begin{center}
\includegraphics[width=7cm]{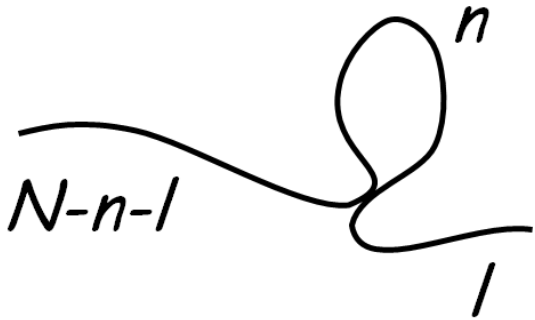}
\end{center}
\caption{A single loop of $n$ monomers created by a sticker at a distance $l$
from one of the ends.}
\label{Figure1}
\end{figure}

Let us first examine the case of an isolated chain with a single sliding
sticker binding two Kuhn segments, as shown in Figure \ref{Figure1}. In this
figure, one can identify one crosslink and two chain ends, which divide the
chain into three threads. If the chain has a total length $N$, the first
thread is a tail of size $l$, the second is a loop of size $n,$ and then
another tail of length $N-n-l$.

\begin{table}[tbp]
\caption{Numerical values of the critical exponent $\protect\nu $ and star
exponents $\protect\sigma$, obtained by interpolation of Monte Carlo results
for star polymers \protect\cite{Grassberger} in $d=3$ dimensions.}
\label{Table}%
\begin{tabular*}{\hsize}{@{\extracolsep{\fill}}r|r|r|r|r|r|r|r|r|r}
$d\nu $ & $\sigma _{1}$ & $\sigma _{2}$ & $\sigma _{3}$ & $\sigma _{4}$ & $%
\sigma _{5}$ & $\sigma _{6}$ & $\sigma _{7}$ & $\sigma _{8}$ & $\sigma _{9}$
\\ \hline
$1.776$ & $0.079$ & $0$ & $-0.193$ & $-0.479$ & $-0.849$ & $-1.292$ & $%
-1.803 $ & $-2.377$ & $-3.014$ \\ \hline\hline
$\sigma _{10}$ & $\sigma _{11}$ & $\sigma _{12}$ & $\sigma _{13}$ & $\sigma
_{14}$ & $\sigma _{15}$ & $\sigma _{16}$ & $\sigma _{17}$ & $\sigma _{18}$ &
$\sigma _{19}$ \\ \hline
$-3.709$ & $-4.449$ & $-5.240$ & $-6.084$ & $-6.975$ & $-7.916$ & $-8.898$ &
$-9.926$ & $-10.991$ & $-12.093$%
\end{tabular*}%
\end{table}

The partition function of the structure sketched in Figure \ref{Figure1},
has a general form $Z\sim s^{N}N^{\gamma _{\mathcal{L}%
}-1}g(l/N,n/N,(N-l-n)/N)$, where the exponent is given by eq. (\ref{genexpr}
), with $q_{loop}=1$, with two vertices of one leg $\sigma _{1}$, and one
vertex of four legs $\sigma _{4},$ i.e. $\gamma _{\mathcal{L}}-1=2\sigma
_{1}+\sigma _{4}-d\nu $ (see Table \ref{Table} for numerical values).
Without the explicit knowledge of the crossover function $g$, the exact
partition function for the system of Figure \ref{Figure1}, with the sliding
degree of freedom, cannot be obtained. However the scaling behavior can be
derived from the singular form of $g$ in the appropriate limit, obtained by
making some of its arguments going to $0$ or to $1$, where $g$ is expected
to behave either as a power law or as a constant. With this in mind, we
propose a power law form of the partition function that will permit us to
properly interpolate between the singular behavior of the crossover function
$g$ in the different limits, i.e.

\begin{equation}
Z\sim n^{x}l^{y}\left( N-n-l\right) ^{z}  \label{freeloop}
\end{equation}
where the exponents $x,y,z$ can be fixed from the known limits. In this
simple case, one of these limits is, for example, $l/N\rightarrow 0$, $%
n/N\sim 1$, and $(N-n-l)/N\sim 1$. Our procedure will become apparent in
what follows.

The partition function of the system requires the integration over the
sliding degree of freedom. The final result has a general structure of the
form

\begin{equation}
Z(N,m)\sim e^{-m \varepsilon /kT}s^{N}\tilde{Z}(N,m)
\end{equation}
where $m$ is the number of stuck colloids, and the factor $e^{-m \varepsilon
/kT}$ takes into account the interaction energy in the sticking process. The
entropic part contains the trivial factor $s^{N}$depending only on the
number of monomers contained in the polymeric network. $\tilde{Z}$ takes
into account the nontrivial contribution to the configurational part of the
partition function which depends on the topology of the polymeric network,
but also on that induced by the presence of $m$ colloids adhered to the
polymer. From now on, we will concentrate on $\tilde{Z}$, and omit the
tilde, for the ease of notation, where confusion could not occur.

For the case of one polymer chain and one colloid, the nontrivial part of
the partition function takes the form

\begin{equation}
Z(N,m=1)\sim 2\int_{0}^{N/2}dl\int_{0}^{N-2l}dn\,g(l/N,n/N,(N-l-n)/N)
\label{ZNm}
\end{equation}
The integration limits take into account that the system is invariant under
the permutation of the two tails and therefore we can restrain the
integration domain by the line $l=(N-n)/2$, which implies that the tail of
size $l$ is the smallest tail. We have to further split the integration
domain into separate regions of different topologies due to the fact that
each topology requires a set of exponents $x,y,z$ in equation (\ref{freeloop}
) to consistently interpolate between the appropriate limits. We can argue
that these regions are characterized by I) small loop, $n<l<N-l-n$, II)
intermediate loop, $l<n<N-l-n$; and III) big loop, $l<N-l-n<n$. Due to the
fact that we take $l$ as the smallest tail without loss of generality. We
when have

\begin{eqnarray}
Z(N,m &=&1)\sim \int_{0}^{N/3}dl\int_{0}^{l}dnn^{x}l^{y}\left( N-n-l\right)
^{z}+\int_{N/3}^{N/2}dl\int_{0}^{N-2l}dnn^{x}l^{y}\left( N-n-l\right) ^{z}
\notag \\
&+&C_{1}\int_{0}^{N/3}dn\int_{0}^{n}dln^{x^{\prime }}l^{y^{\prime }}\left(
N-n-l\right) ^{z^{\prime }}\,  \notag \\
&&+C_{1}\int_{N/3}^{N/2}dn\int_{0}^{N-2n}dln^{x^{\prime }}l^{y^{\prime
}}\left( N-n-l\right) ^{z^{\prime }}  \notag \\
&+&C_{2}\int_{N/3}^{N/2}dn\int_{N-2n}^{(N-n)/2}dln^{x^{\prime \prime
}}l^{y^{\prime \prime }}\left( N-n-l\right) ^{z^{\prime \prime }}  \notag \\
&&+C_{2}\int_{N/2}^{N}dn\int_{0}^{(N-n)/2}dln^{x^{\prime \prime
}}l^{y^{\prime \prime }}\left( N-n-l\right) ^{z^{\prime \prime }}  \label{Z}
\end{eqnarray}%
where the constants $C_{1}$ and $C_{2}$ measure the relative weight between
the functional form taken for each of the contributions. In what follows we
will determine the three sets of exponents and evaluate the contributions of
each term to the partition function.

II) Intermediate loop, $l<n$. In this region, only three limits are
possible. In the first limit we consider that all legs and the loop are of a
similar size and of order $N$. Therefore, according to eq. (\ref{freeloop}),
$Z\sim N^{x^{\prime }+y^{\prime }+z^{\prime }}$. Then, making use of the
general expression for the topological exponents given in equation (\ref%
{genexpr}), one can write

\begin{equation}
x^{\prime }+y^{\prime }+z^{\prime }=2\sigma _{1}+\sigma _{4}-d\nu
\label{xyz}
\end{equation}
corresponding to a topology of two free ends at the extremes of the two
tails, plus a vertex of four legs where the loop closes in the middle of the
polymer. In the second limit, if the small tail vanishes, $l\rightarrow 1$ ($%
l/N\rightarrow 0$) with $n\sim N/2$ and $(N-n-1)\sim N/2$, the topology of
the system evolves towards a long tail with the loop at the end,
characterized by a topology with the presence of one free end at the extreme
of the remaining long tail, and a vertex of three legs for the loop closed
at the other end of the tail. Then $Z\sim n^{x^{\prime }}1^{y^{\prime
}}(N-n)^{z^{\prime }}$. Since this expression has to be compatible with the
known singular behavior of the system with the mentioned topology (\ref%
{genexpr}), we further have

\begin{equation}
x^{\prime }+z^{\prime }=\sigma _{1}+\sigma _{3}-d\nu
\end{equation}
Finally, in the third limit, if both, the small tail and the loop, vanish, $%
l<n\rightarrow 1$, we get a free chain topology and thus

\begin{equation}
z^{\prime }=2\sigma _{1}
\end{equation}
These three equations permit us to find the numerical value of the desired
exponents for the interpolating function that will be used in the
integration to derive the singular behavior of the partition function. We
obtain $x^{\prime }=$ $\sigma _{3}-\sigma _{1}-d\nu $, $y^{\prime }=\sigma
_{1}+\sigma _{4}-\sigma _{3}$, and $z^{\prime }=2\sigma _{1}$.

We can now write the partition function of this domain in the form of nested
integrals

\begin{eqnarray}
Z_{II} &\sim &\int_{0}^{N/3}dnn^{\sigma _{3}-\sigma _{1}-d\nu
}\int_{0}^{n}dll^{\sigma _{1}+\sigma _{4}-\sigma _{3}}(N-n-l)^{2\sigma _{1}}+
\notag \\
&&+\int_{N/3}^{N/2}dnn^{\sigma _{3}-\sigma _{1}-d\nu
}\int_{0}^{N-2n}dll^{\sigma _{1}+\sigma _{4}-\sigma _{3}}\left( N-n-l\right)
^{2\sigma _{1}}  \label{Zintfreeloop}
\end{eqnarray}
In both terms, since $\sigma _{1}+\sigma _{4}-\sigma _{3}\approx -0.2>-1$
and $2\sigma _{1}=0.16>1$ \cite{GrassbergerDNA}, the inner integral
converges. However, the inner integral in the first contribution behaves as $%
n^{\sigma _{1}+\sigma _{4}-\sigma _{3}+1}$ for small $n$. Therefore the
outer integral is diverging due to lower bound, leading to a contribution of
the order $\Delta ^{\sigma _{4}-d\nu +2}N^{2\sigma _{1}}\approx \Delta
^{\sigma _{4}-d\nu +2}N^{0.16}$, where $\Delta $ is the lower cutoff for the
size of the loop, of the order of one monomer. The second contribution is
convergent, yielding $N^{\sigma _{4}-d\nu +2\sigma
_{1}+2}\int_{1/3}^{1/2}dtt^{\sigma _{3}-\sigma _{1}-d\nu
}\int_{0}^{1-2t}dss^{\sigma _{1}+\sigma _{4}-\sigma _{3}}\left( 1-t-s\right)
^{2\sigma _{1}}=0.31N^{\sigma _{4}-d\nu +2\sigma _{1}+2}\approx
0.31N^{-0.098}$. Therefore, in the limit $N\rightarrow \infty $ the
partition function is dominated by the first term

\begin{equation}
Z_{II}\sim \Delta ^{\sigma _{4}-d\nu +2}N^{2\sigma _{1}}  \label{Zsmalltail}
\end{equation}
This result indicates that big loops are not entropically favorable and that
the system will tend to reduce the size of the loop to the minimum.
Moreover, since in this limit the small tail is, by construction, smaller
than the loop, the most probable conformation is hence a minimal loop at the
end of a long tail.

I) To evaluate $Z_{I}$, one has to consider that the loop is smaller than
the shortest tail, $n<l$. When the sizes of the loop and the tails are
comparable, we have the same scaling as in the II-part. Hence, in the latter
case

\begin{equation}
x+y+z=2\sigma _{1}+\sigma _{4}-d\nu
\end{equation}
Notice that when the loop and the tails are comparable and of the order of $%
N $, the integrands of the II- and I- part should be comparable. Since the
functional form is also the same for both cases, we have that the constant $%
C_{1}\sim 1$.

As before, to fix the exponents we identify the singular limits compatible
with the condition $n<l$. Taking $n\rightarrow 1$, the free chain limit is
recovered and hence

\begin{equation}
y+z=2\sigma _{1}
\end{equation}
Furthermore, when the small tail vanishes, this implies a simultaneous
vanishing of both the loop and the tail, $n<l\rightarrow 1$. One gets

\begin{equation}
z=2\sigma _{1}
\end{equation}
Finally, one also obtains $x=\sigma _{4}-d\nu $, $y=0$. The partition
function reads

\begin{equation}
Z_{I}\sim \int_{0}^{N/3}dl\int_{0}^{l}dnn^{\sigma _{4}-d\nu
}(N-n-l)^{2\sigma _{1}}+\int_{N/3}^{N/2}dl\int_{0}^{N-2l}dnn^{\sigma
_{4}-d\nu }\left( N-n-l\right) ^{2\sigma _{1}}
\end{equation}
Since again the integrand of the inner integrals diverge the integral is
dominated by $n\rightarrow \Delta $. The outer integrals have a regular
integrand and the whole term scales as $N^{2\sigma _{1}+1}$. Then the final
expression for the partition function of this part is

\begin{equation}
Z_{I}\sim \Delta ^{\sigma _{4}-d\nu +1}N^{2\sigma _{1}+1}  \label{Zfreeloop}
\end{equation}
This partition function indicates that under the condition $n<l$
the most favorable conformation is the existence of a minimal loop
traveling along the chain. Notice that this contribution has an
extra $N$ factor as compared to the case $II$, that is
$Z_{I}/Z_{II}\sim N/\Delta \rightarrow \infty $, for very long
chains, indicating that this contribution will be dominant.
Similar conclusion is drawn for the case knots in chains
\cite{Ben-Naim}.

III) The third contribution in (\ref{Z}) is obtained as before. For this
case the relevant limits are: both tails and the loop are of the same order,
one tail vanishes, $l\rightarrow 1$, both tails tend to vanish $%
l<N-l-n\rightarrow 1$. With these limits we find $x^{\prime \prime }=-d\nu $
, $y^{\prime \prime }=\sigma _{1}+\sigma _{4}-\sigma _{3}$, and $z^{\prime
\prime }=\sigma _{1}+\sigma _{3}$. Furthermore, since $Z_{III}$ should be
comparable with $Z_{I}$ and $Z_{II}$ when the two tails are of the same
order we conclude that $C_{2}$ is of order $1$.

\begin{eqnarray}
Z_{III} &\sim &\int_{N/2}^{N}dn\int_{0}^{(N-n)/2}dln^{-d\nu }l^{\sigma
_{1}+\sigma _{4}-\sigma _{3}}\left( N-n-l\right) ^{\sigma _{1}+\sigma _{3}}+
\notag \\
&&\int_{N/3}^{N/2}dn\int_{N-2n}^{(N-n)/2}dln^{-d\nu }l^{\sigma _{1}+\sigma
_{4}-\sigma _{3}}\left( N-n-l\right) ^{\sigma _{1}+\sigma _{3}}
\end{eqnarray}

Both integrands have integrable divergences and, therefore, the $N$
dependence of $Z_{III}$ can be trivially obtained as before $N^{\sigma
_{4}-d\nu +2\sigma _{1}+2}\left[ \int_{1/2}^{1}dtt^{-d\nu
}\int_{0}^{(1-t)/2}dss^{\sigma _{1}+\sigma _{4}-\sigma _{3}}\left(
1-t-s\right) ^{\sigma _{1}+\sigma _{3}}\right.$ $\left.+
\int_{1/3}^{1/2}dtt^{-d\nu }\int_{1-2t}^{(1-t)/2}dss^{\sigma _{1}+\sigma
_{4}-\sigma _{3}}\left( 1-t-s\right) ^{\sigma _{1}+\sigma _{3}}\right]$ $%
\sim 0.44N^{\sigma _{4}-d\nu +2\sigma _{1}+2}$

We can then conclude that the dominant contribution to the partition
function of a sticker in a free chain is given by $Z\sim Z_{I}\sim \Delta
^{\sigma _{4}-d\nu +1}N^{2\sigma _{1}+1}$, since $Z_{II}/Z_{I}\sim N^{-1}$
and $Z_{III}/Z_{I}\sim N^{\sigma _{4}-d\nu +1}/\Delta ^{\sigma _{4}-d\nu
+1}\sim N^{-1.26}$.

Therefore the overall $N$-dependence of the partition function is that of a
bare chain with an extra power of $N$ that takes into account the freedom of
the location of the small loop along the chain. The loop cut-off size $%
\Delta $ could be linked to the local stiffness of the chain (persistence
length). The optimal configuration is a small loop freely sliding along a
linear chain. This is a general result and, as we describe later, is
applicable to the loops sliding in the corona of a micelle or the arm of a
star polymer. The position of the loop $l$ does not enter the final
expression and the distribution of positions of the small loop along the
chain and hence the probability of finding the loop (the sticker) at a
position $l$ along the chain backbone $P(l)$ is uniform up to a distance $%
\Delta $ from the chain ends.

\begin{figure}[ht]
\begin{center}
\includegraphics[width=8.5cm]{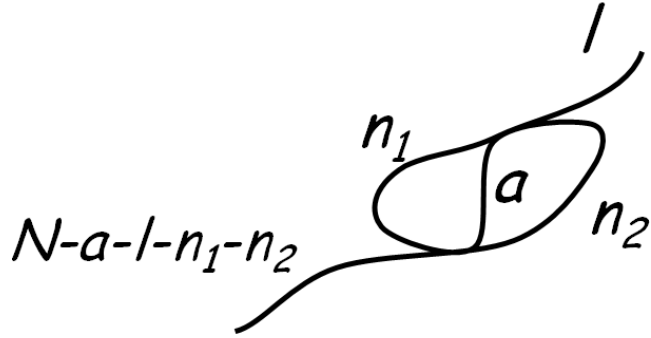}
\end{center}
\caption{{Ailed loop created by two stickers on the same chain.}}
\label{fig2}
\end{figure}

To end this section, let us consider the case of $m$ stickers in one linear
chain to the light of the results that we have just derived. As we have
seen, the more important contribution to the partition function is the
degree of freedom of each small loop along the chain backbone. This
contribution is lost if the loop becomes large as well as in conformations
in which loop-like structures are linked by a sticker. For example, in ailed
loops (Figure \ref{fig2}) the translational entropy gives a factor $N$
while, if the loops were independent, the entropy would be of order $N^{2}$
(see Appendix \ref{secailedloop}). Then the partition function for a system
of $m$ stickers in a free chain can be straightforwardly generalized
considering the independent contributions of the individual loops, provided
that $m\Delta \ll N$. Thus, from the expression (\ref{Zfreeloop}) one obtains

\begin{equation}
Z\sim \Delta ^{m(\sigma _{4}-d\nu +1)}N^{m}N^{2\sigma _{1}}  \label{mloops}
\end{equation}

Notice that the factor $N^{2\sigma _{1}}$ stands for the entropy of the
supporting chain. The result given in eq. (\ref{mloops}) is only valid in
the case that the sticker-sticker interactions are negligible and therefore,
when the chain is not saturated.

\subsection{Stickers in star polymers\label{secstickers}}

In the following, let us assume that the star polymer has $p$ arms, each of
length $N$. We will proceed by analyzing the dominant diagrams starting with
one colloid and then generalizing the result to an arbitrary number $m$ of
stickers adhered to the arms of the polymer as before.

\subsubsection{One loop on one arm}

\begin{figure}[ht]
\begin{center}
\includegraphics[width=5cm]{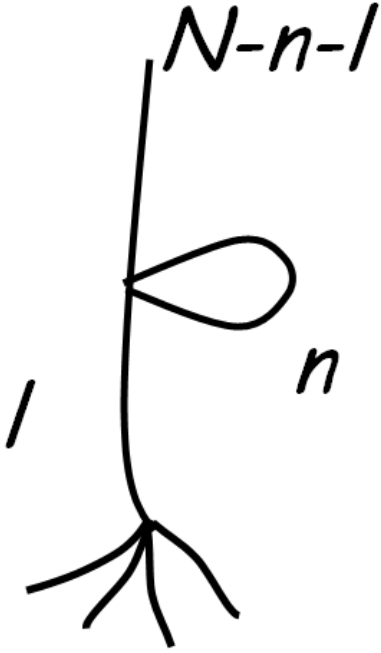}
\end{center}
\caption{{Sliding loop of size $n$ on the arm of a micelle at the distance $%
l $ from the center.}}
\label{fig7}
\end{figure}

The simplest case is that of one loop in one of the arms of the star
polymer, as shown in Figure \ref{fig7}. The main difference with the respect
to previous analysis is here the presence of the topological constraints at
the vertex where the $p$ arms meet, giving strong excluded volume
interactions in the vicinity. This is reflected in the partition function of
the system by the presence of highly negative exponent $\sigma _{p}$
corresponding to strong excluded volume effects in the center.

Denoting the size of the loop by $n$ and the distance of the loop from the
center by $l$, we write the partition function of the system star polymer --
colloid under the interpolating form already used $Z\sim
n^{x}l^{y}(N-n-l)^{z}$, where $x$, $y$ and $z$ are exponents to be
determined by matching the appropriate limits. Furthermore, as before, the
set of exponents is not unique but depends on the relative sizes of the
three entities in which we have divided the chain, i.e. loop, tail, and
distance to the vertex. We demand that these different expressions for the
interpolating function continuously crossover at the limits of their
validity.

Following the same analysis of the previous section, the interpolating
function have to be split into six different sets of exponents due to lack
of symmetry of this case. However, we know that the partition function is
going to be dominated by a small loop traveling along the arm. To simplify
our analysis we will concentrate on the scaling form of this regime, defined
by a set of exponents in the region $n<l<N-n-l$, and a different set in the
region $n<N-n-l<l$. After analyzing all the corresponding limits, one has

\begin{eqnarray}
Z &\sim &\int_{0}^{N/3}dl\int_{0}^{l}dnn^{\sigma _{4}-d\nu }(N-n-l)^{\sigma
_{p}+p\sigma _{1}}+\int_{N/3}^{N/2}dl\int_{0}^{N-2l}dnn^{\sigma _{4}-d\nu
}(N-n-l)^{\sigma _{p}+p\sigma _{1}}  \notag \\
&&+\int_{N/3}^{N/2}dl\int_{N-2l}^{(N-l)/2}dnn^{\sigma _{4}-d\nu }l^{\sigma
_{p}+p\sigma _{1}}+\int_{N/2}^{N}dl\int_{0}^{(N-l)/2}dnn^{\sigma _{4}-d\nu
}l^{\sigma _{p}+p\sigma _{1}}  \label{Zloopstar}
\end{eqnarray}

As expected, the inner integrals diverge for this small loop case when $%
n\rightarrow \Delta $, which reflects the aforementioned dominance of small
loops with respect to large loops. After integration we get that the
dominant contribution scales as

\begin{equation}
Z\sim \Delta ^{\sigma _{4}-d\nu +1}N^{\sigma _{p}+p\sigma _{1}}N
\label{oneloop}
\end{equation}
The term $N^{\sigma _{p}+p\sigma _{1}}$ is the partition function of a star
polymer of $p$ arms, while the additional $N$ factor stands for the
translational entropy of the sliding small loop. The other topologies that
we have not explicitly analyzed imply either a loop of order $N$ or small
loops that are interacting with the vertices. In all these cases the
translational entropy of the loop is lost and these contributions are
therefore subdominant. The integrand in the partition function (\ref%
{Zloopstar}), upon integration over $n$, allows us to derive the
distribution of these small loops along the chain, giving

\begin{equation}
P(l)\sim \left\{
\begin{array}{c}
(N-l)^{\sigma _{p}+p\sigma _{1}},0<l<N/2 \\
l^{\sigma _{p}+p\sigma _{1}},N/2<l<N%
\end{array}
\right.  \label{Pl}
\end{equation}
which indicates a slight repulsion of the small loop from the center of the
star.

\subsubsection{Several loops on the same arm}

The partition function of several loops on the same arm can be obtained by a
straightforward generalization of the previous result, as we have done for
the single chain case. We obtain

\begin{equation}
Z\sim \Delta ^{m(\sigma _{4}-d\nu +1)}N^{m}N^{\sigma _{p}+p\sigma _{1}}
\label{dominant_diagram}
\end{equation}
which basically contains the bare entropy of the star polymer together with $%
N^{m}$ factor due to the translational entropy of $m$ loops. We have
implicitly considered that the number of monomers in the $i^{th}$-branch, $%
n_{i}$ is of order $N$. This result is however limited to the case $m\Delta
\ll N$, due to the fact that no interactions between loops has been
considered.

This result can be further generalized to the case where the $m$ stickers
are distributed among $p$ branches. Let us consider a star polymer with $p$
-arms of $n_{1},n_{2},\dots ,n_{p}$ monomers each, with $%
\sum_{i=1}^{p}n_{i}=N$, where colloidal particles can adhere from a given
bulk solution. The complete partition function of $m$ colloids attached to
this polymer is given by
\begin{equation}
Z(m)\propto \sum_{m_{1}+m_{2}+\dots +m_{p}=m}e^{-m\varepsilon
/kT}s^{N}\Delta ^{m(\sigma _{4}-d\nu +1)}N^{m}\left( \frac{n_{1}}{N}\right)
^{m_{1}}\left( \frac{n_{2}}{N}\right) ^{m_{2}}\dots \left( \frac{n_{p}}{N}
\right) ^{m_{p}}N^{\sigma _{p}+p\sigma _{1}}  \label{star_partition_0}
\end{equation}
where use has been made of eq. (\ref{dominant_diagram}) to introduce the
dominant contribution to the entropy of the system, and $%
\sum_{i=1}^{p}m_{i}=m$. Branches are assumed to be discernable due to
polydispersity considerations. However, if the size of the branches $n_{i}$
are comparable and $n_{i}\sim N$ $n_{1}/N\simeq n_{2}/N\simeq \ldots \simeq
n_{p}/N$ we can approximately write

\begin{equation}
Z(m)\propto \frac{(m+p-1)!}{m!(p-1)!}e^{-m\varepsilon /kT}s^{N}\Delta
^{m(\sigma _{4}-d\nu +1)}\left( \frac{N}{p}\right) ^{m}N^{\sigma
_{p}+p\sigma _{1}}  \label{star_partition}
\end{equation}

The combinatorial factor stands for all combinations of $m_{1},m_{2},\ldots
,m_{p}$ such that its sum gives $m$.

\subsubsection{m stickers looping in many arms}

\begin{figure}[ht]
\begin{center}
\includegraphics[width=4cm]{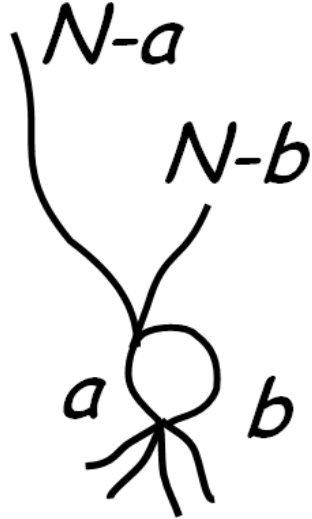}
\end{center}
\caption{{A loop between two arms of a micelle created by one sticker in the
center of the micelle.}}
\label{fig5}
\end{figure}

The exact consideration of the general case is a formidable task in view of
the multiplicity of exponents and integration domains that we have shown in
the simple cases analyzed above. However, we prove here that when two arms
are connected by one sticker, there is a significant entropy reduction with
respect to the case where the sticker form one loop in one arm. Therefore,
the dominant configuration for the general case is that in which the $m$
stickers form small traveling loops in individual arms. To prove this, let
us first consider a loop formed by two different arms as shown in Figure \ref%
{fig5}. The sticker splits a first chain into a segment $a$ close to the
center and a free tail $N-a$. The second chain is split into a segment $b$
and a free tail $N-b$, while the loop size is $n=a+b$. Since two arms are
symmetrical we sort the tails by their respective lengths, $a<b$ this leads
to $N-b<N-a$. The general form of the partition function is

\begin{equation}
Z \sim a^{x_{1}}a^{x_{2}}(N-a)^{z_{1}}(N-b)^{z_{2}}  \label{clue}
\end{equation}

where the exponents $x_{1}$, $x_{2}$, $z_{1}$, $z_{2}$ are fixed by the
limits. The dominant contribution to the partition function is found in the
limit $a<b<N-b<N-a$. This can be intuitively understood by noticing that the
size of the loop is directly related to the location of the sticker. In this
way, when the sticker is located near the chain end (a large loop) there is
a strong entropy reduction due to the reduction of the two tails to create a
big loop. Hence, the dominant contribution is that of a small loop near the
core of the star. This argument also indicates that the translational
entropy of the small loop discussed in the previous cases is here lost.
Therefore the dominant contribution to the partition function reads

\begin{equation}
Z\sim \int_{0}^{N/2}dbb^{\sigma _{p+2}-\sigma _{p}-d\nu
}\int_{0}^{b}daa^{\sigma _{p}-\sigma _{p+2}+\sigma
_{4}}(N-a)^{z_{1}}(N-b)^{z_{2}}  \label{Zcenloop}
\end{equation}

where use has been made of the relevant limits to fix the exponents, with $%
z_{1}+z_{2}=\sigma _{p}+p\sigma _{1}$ as imposed by the limit $a$,$%
b\rightarrow 0$. Since $\sigma _{p}-\sigma _{p+2}+\sigma _{4}>-1$, the inner
integral converges, giving, $Z\sim \int_{0}^{N/2}db \, b^{\sigma _{4}-d\nu
+1}N^{\sigma _{p}+p\sigma _{1}}$, while this integral diverges on the lower
limit $\Delta $, since $\sigma _{p+2}-\sigma _{p}-d\nu <-1$ for all $p$. One
finally obtains

\begin{equation}
Z\sim \Delta ^{\sigma _{4}-d\nu +2}N^{\sigma _{p}+p\sigma _{1}}
\label{centerloop}
\end{equation}

\begin{figure}[ht]
\begin{center}
\includegraphics[width=6cm]{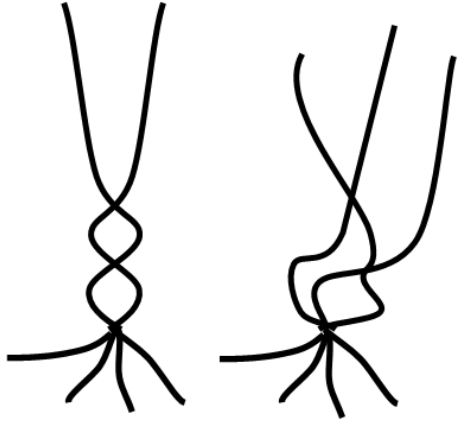}
\end{center}
\caption{{Two examples of nested loops formed by two stickers in the arm of
a micelle.}}
\label{fig6}
\end{figure}

Similar arguments lead to the conclusion that a larger number of loops
between two different arms or nested loops of several arms as shown in
Figure \ref{fig6} will degenerate into immobile loop in the center: lower
limit exponent $<-1$ for any $p$, including $p=4$, and the nested integral
in (\ref{Zcenloop}) is of order of $b$ regardless the integrated function.

%\newpage

\section{Equilibrium between colloidal stickers and a $p$-armed star \label%
{seceq}}

With the help of the results obtained so far we analyze here the equilibrium
between a solution of small colloidal particles and polymeric structures.
This method can be applied to determine the partition function of a system
in which small colloids can adhere to polymers of complex topology. On one
hand, it is useful to find how the topological constrains can modify the
preferred location of the colloids in the system, but also how colloids can
induce the binding of different branches. Therefore, it is not a priori
evident the final equilibrium structure.

In the first place, we study the case of a star polymer of $p$-arms in
equilibrium with a suspension of colloids. In the second place, we will
drive our attention to the case of block copolymer micelles, where the
equilibrium can change the properties of the micelle as such. This latter
case is of interest in the moulding of porous materials from the
self-assembly of surfactants through silica colloids associating with the
corona of copolymeric micelles, as it is the case for porous materials of
type SBA or MCM, for example.

Making use of eqs. (\ref{star_partition}), the grand partition function for
the system then reads
\begin{equation}
\Xi \propto \sum_{m=0}^{\infty }e^{\mu m/kT}Z(m)\simeq s^{pL}\left(
pL\right) ^{\sigma _{p}+p\sigma _{1}}\sum_{m=0}^{\infty }\left[ e^{\frac{\mu
-\varepsilon }{kT}}\Delta ^{\sigma _{4}-\nu d+1}L\right] ^{m}\frac{(m+p-1)!}{%
m!(p-1)!}  \label{grand_canonical}
\end{equation}%
Stirling approximation can be used if $m>>1$, while $p$ can be of order $1$,
thus the combinatorial factor turning into $m^{p-1}/(p-1)!$. Furthermore, we
have introduced the number of monomers per arm, $L$, so that $N=pL$. The
grand-canonical potential $\Omega $ is defined as usual $\Omega =-kT\ln \Xi $%
. Defining $Y(\mu )\equiv e^{\frac{\mu -\varepsilon }{kT}}\Delta ^{\sigma
_{4}-\nu d+1}L$ one finds that the sum is converging always that $Y(\mu )<1$%
. In particular, in the limit $Y(\mu )\rightarrow 1$ we find
\begin{equation}
\Xi \propto \frac{1}{(1-Y(\mu ))^{p}}  \label{grand_canonical_asymp}
\end{equation}%
Therefore, colloids saturate the star polymer when the chemical potential
rises the critical value $\mu _{c}=\varepsilon +kT\ln \left( \Delta ^{\sigma
_{4}-\nu d+1}L\right) $. In particular, we find that the average number of
colloids that aggregate in the star is given by
\begin{equation}
\langle m\rangle =\frac{\sum_{m=0}^{\infty }Y(\mu )^{m}m^{p}}{%
\sum_{m=0}^{\infty }Y(\mu )^{m}m^{p-1}}\rightarrow \frac{1}{1-Y(\mu )}\frac{%
\Gamma (p+1)+\zeta (-p)}{\Gamma (p)+\zeta (1-p)}\simeq \frac{p}{1-Y(\mu )}
\label{average}
\end{equation}%
where the last expression corresponds to the limit $Y\rightarrow 1^{-}$ and $%
p$ is taken $p\geq 4$. Therefore, above the critical value of the chemical
potential the star saturates and $m$ becomes of the order of $1/\alpha \gg 1$%
, where $\alpha $ is $\Delta /N=\Delta /pL$. The quantity $\alpha m$ can be
regarded as the fraction of polymer wrapping colloids, which tend to $1$ at
saturation.

The analysis of the saturated state cannot be done with the same rigorous
approach developed so far. We will assume that saturation occurs because of
the limitation of the available polymer for the colloids in one arm due to
the presence of other colloids in the same arm, in the spirit of van der
Waals equation of state. One can then approximately write $(L-m\Delta /p)^{m}
$ instead of $L^{m}$ and $(L-m\Delta /p)^{\sigma _{p}+p\sigma _{1}}$in eq. (%
\ref{star_partition}), to account for the limitation of the size of the
polymeric threads at fixed number $m$ of minimal loops. This approximation
will be valid always that $m\Delta \ll N$. One can finally write
\begin{equation}
Z(m,p)\propto \frac{m^{p-1}}{(p-1)!}\,e^{-m\varepsilon /kT}s^{pL}\Delta
^{m(\sigma _{4}-d\nu +1)}L^{m}(pL)^{\sigma _{p}+p\sigma _{1}}\left( 1-m\frac{%
\Delta }{pL}\right) ^{m}\left( 1-m\frac{\Delta }{pL}\right) ^{\sigma
_{p}+p\sigma _{1}}  \label{star_partition_sat}
\end{equation}%
The partition function can then be written as
\begin{equation}
\Xi \propto \sum_{m=0}^{N/\Delta }Y(\mu )^{m}\left( 1-\frac{m\Delta }{pL}%
\right) ^{m+\sigma _{p}+p\sigma _{1}}m^{p-1}
\end{equation}%
Let us center our attention on the value of the sum
\begin{equation}
I_{p-1}\equiv \sum_{m=0}^{1/\alpha }Y(\mu )^{m}\left( 1-\alpha m\right)
^{m+\sigma _{p}+p\sigma _{1}}m^{p-1}  \label{part_sat}
\end{equation}%
in the limit $\alpha \rightarrow 0$, near the transition, $Y\rightarrow
1^{\pm }$. Then, the summand can be rewritten as
\begin{equation}
Y(\mu )^{m}\left( 1-\alpha m\right) ^{m+\sigma _{p}+p\sigma
_{1}}m^{p-1}\simeq e^{m\ln Y+(m+\sigma _{p}+p\sigma _{1})\ln (1-\alpha
m)+(p-1)\ln m}  \label{aprox}
\end{equation}%
The exponent in the second expression can be regarded as an effective
Hamiltonian
\begin{equation}
H(m)\equiv -\left( m\ln Y+(m+\sigma _{p}+p\sigma _{1})\ln (1-\alpha
m)+(p-1)\ln m\right)   \label{hamiltonian}
\end{equation}%
To analyze the stability of the system, let us calculate the saddle point $%
m^{\ast }$ by demanding that
\begin{equation}
\left. \frac{\partial H}{\partial m}\right\vert _{m^{\ast }}=0
\label{saddle}
\end{equation}%
This gives the so-called equation of state, which reads in our case
\begin{equation}
\ln Y+\ln (1-\alpha m^{\ast })-\frac{\alpha (m^{\ast }+\sigma _{p}+p\sigma
_{1})}{1-\alpha m^{\ast }}+\frac{p-1}{m^{\ast }}=0  \label{eq_state}
\end{equation}%
Let us assume that near the transition, $\ln Y\ll 1$, the fraction of
wrapped polymer $\alpha m^{\ast }$ is small. Then, from eq. (\ref{eq_state})
we obtain an estimate of the average value of $m^{\ast }$. To the lowest
order in $\alpha $ we find $\ln Y-2\alpha m^{\ast }-\alpha (\sigma
_{p}+p\sigma _{1})+(p-1)/m^{\ast }=0$, which gives
\begin{equation}
m^{\ast }=\frac{\ln Y-\alpha (\sigma _{p}+p\sigma _{1})}{4\alpha }\pm \frac{1%
}{4\alpha }\sqrt{8\alpha (p-1)+(\ln Y-\alpha (\sigma _{p}+p\sigma _{1}))^{2}}
\label{m}
\end{equation}%
The negative sign gives an unphysical $m^{\ast }<0$ solution. This
expression gives a crossover between values of $Y$ smaller than 1 and values
larger than 1 for finite $\alpha $. The assymptotic behavior is obtained by
expanding the expression for small $\alpha $, in fact for $\ln Y/\sqrt{%
\alpha }$ larger than 1. At the crossover, $|\ln Y|\sim \alpha ^{1/2}$ .
Therefore, if $1>|\ln Y|>\alpha ^{1/2}$ the solution reads
\begin{eqnarray}
m_{+}^{\ast } &\simeq &\frac{\ln Y}{2\alpha }+\frac{p-1}{\ln Y}-\frac{%
2(p-1)^{2}}{\ln ^{3}Y}\alpha +\mathcal{O}\left( \alpha \ln Y\right) \hspace{%
0.5cm}\mbox{for}\;\;Y>1  \label{plus_large} \\
m_{-}^{\ast } &\simeq &\frac{p-1}{|\ln Y|}-\frac{2(p-1)^{2}}{|\ln ^{3}Y|}%
\alpha +\mathcal{O}\left( \alpha \ln Y\right) \hspace{0.5cm}\mbox{for}\;\;Y<1
\label{minus_large}
\end{eqnarray}%
For values of the control parameter $\alpha ^{1/2}>|\ln Y|\geq 0$, the
appropriate expansion reads for both branches
\begin{equation}
m^{\ast }\simeq \sqrt{\frac{p-1}{2\alpha }}-\frac{\sigma _{p}+p\sigma _{1}}{4%
}+\frac{\ln Y}{4\alpha }+\mathcal{O}\left( \alpha ^{1/2},\frac{\ln Y}{\alpha
^{1/2}}\right)   \label{plus_small}
\end{equation}%
This result indicates that the free energy is analytic through the
transition and the saturation is therefore a crossover process as the
chemical potential increases for finite $\alpha $, due to the existence of
this crossover regime $\alpha ^{1/2}>|\ln Y|\geq 0$. However, the range of
validity of eq. (\ref{plus_small}) decreases as $\alpha \rightarrow 0$
indicating that the behavior of $m^{\ast }$ will become singular precisely
in this limit. To further analyze the character of such singular behavior,
we have to consider the limit $\ln Y\rightarrow 0$ with $|\ln Y|>\alpha
^{1/2}$. The appropriate order parameter in this singular limit is the
wrapped fraction of polymer, $\psi \equiv \alpha m^{\ast }$. Since for $\psi
=\alpha m_{-}^{\ast }=0$ for $Y<1$ while $\psi =\alpha m_{+}^{\ast }\sim \ln
Y$ if $\alpha \rightarrow 0$. In addition, notice that $\alpha m^{\ast }\sim
\alpha ^{1/2}\rightarrow 0$ in the crossover regime, eq. (\ref{plus_small}).
Let us further define $\zeta \equiv \ln Y\ll 1$. We then rewrite the
effective Hamiltonian as
\begin{equation}
\alpha H(m)-\alpha (p-1)\ln \alpha \equiv h(\psi )=-\left( \psi \zeta +(\psi
+\alpha (\sigma _{p}+p\sigma _{1}))\ln (1-\psi )+\alpha (p-1)\ln \psi
\right)   \label{hamiltonian_2}
\end{equation}%
Then, the main integral eq. (\ref{part_sat}) reads
\begin{equation}
I_{p-1}\simeq \frac{1}{\alpha ^{p}}\int_{0}^{1}d\psi \,e^{-\frac{h(\psi )}{%
\alpha }}  \label{part_sat_2}
\end{equation}%
The saddle point analysis then yields
\begin{eqnarray}
\psi _{+}^{\ast } &\simeq &\frac{\zeta }{2}+\alpha \frac{p-1}{\zeta }-\alpha
^{2}\frac{2(p-1)^{2}}{\zeta ^{3}}+\mathcal{O}\left( \alpha \zeta \right)
\hspace{0.5cm}\text{for}\;\;\zeta >0  \label{plus_large_2} \\
\psi _{-}^{\ast } &\simeq &\alpha \frac{p-1}{|\zeta |}-\alpha ^{2}\frac{%
2(p-1)^{2}}{|\zeta ^{3}|}+\mathcal{O}\left( \alpha \zeta \right) \rightarrow
0\hspace{0.5cm}\text{for}\;\;\zeta <0  \label{minus_large_2}
\end{eqnarray}%
where we recall that $\zeta >\alpha ^{1/2}$. Furthermore, the mean field
stability of the solution requires
\begin{eqnarray}
\left. \frac{\partial ^{2}h}{\partial \psi ^{2}}\right\vert _{\psi ^{\ast }}
&\equiv &h^{\prime \prime }(\psi ^{\ast })=\frac{2}{1-\psi ^{\ast }}+\frac{%
\psi ^{\ast }+\alpha (\sigma _{p}+p\sigma _{1})}{(1-\psi ^{\ast })^{2}}%
+\alpha \frac{p-1}{\psi ^{\ast \,2}}\simeq   \notag \\
&&\alpha \frac{p-1}{\psi ^{\ast \,2}}+2+\alpha (\sigma _{p}+p\sigma
_{1})+(3+2\alpha (\sigma _{p}+p\sigma _{1}))\psi ^{\ast }+\mathcal{O}\left(
\psi ^{\ast }\right) ^{3}>0  \label{stability_2}
\end{eqnarray}%
where second equality follows considering that we are near the transition
and therefore $\psi \ll 1$. Notice, however, that the limit $\alpha
\rightarrow 0$ has to be taken before $\psi \rightarrow 0$, i.e. before we
approach the transition. Therefore,  Since $h^{\prime \prime }(\psi )>0$ in
all the range of values of $\psi $, the mean field solution is stable.
Therefore,
\begin{equation}
\left. \frac{\partial ^{2}h}{\partial \psi ^{2}}\right\vert _{\psi ^{\ast
}}\approx 2+3\psi ^{\ast }+\mathcal{O}\left( \psi ^{\ast }\right) ^{3}>0
\end{equation}%
which is always positive, indicating the stability of the mean field
solution. Near the transition, the energy is continuous and behaves as
\begin{eqnarray}
h(\psi _{+}^{\ast }) &\sim &-\frac{\zeta ^{2}}{4} \\
h(\psi _{-}^{\ast }) &\sim &-\alpha (p-1)\,\mbox{sign}(\zeta )\;\rightarrow 0
\end{eqnarray}%
The function $h^{\prime \prime }(\psi ^{\ast })$ is also continous at $\zeta
=0$, since
\begin{eqnarray}
h^{\prime \prime }(\psi _{+}^{\ast }) &\sim &2+\frac{3}{2}\zeta >0 \\
h^{\prime \prime }(\psi _{-}^{\ast }) &\sim &2>0
\end{eqnarray}

In conclusion, for $\alpha \rightarrow 0$ at finite $\left\vert \zeta
\right\vert \ll 1$ the system presents a crossover from a \textit{%
non-decorated} state, $\psi _{-}^{\ast }=0$, to a \textit{decorated} state, $%
\psi _{+}^{\ast }\sim \zeta /2$, in which the polymeric threads are
increasingly covered by colloids. However, it is noteworthy that the second
derivative of the saddle point free energy is continous in the transition,
but we observe that the third derivative has indeed a finite jump. Therefore
this behavior can be interpreted as a higher order phase transition in the
sense of the old Ehrenfest classification, but not in the sense of an order
-- disorder, Ising-like transitions.

\section{Micelles of polymers with stickers \label{secmic}}

Colloidal stickers can aggregate in coronas of block copolymer micelles and
modify the equilibrium structure of micelles, which is precisely the
relevant problem for the formation of mesoporous materials through
self-assembly of block copolymers. After the discussion of the saturation
transition, we address this structural problem. With this purpose, let us
first construct the free energy of a micelle with $p$ block copolymers and
with $m$ adhered colloids to consider afterwards the global equilibrium of a
system of many different-sized micelles with free chains and colloids in the
bulk.

Since the colloidal stickers interact only with the hydrophilic
blocks forming the corona, only the corona contribution is
affected by the presence of stickers and the partition function
can be split into two factors, namely, the contribution due to the
corona and that due to the core.

Thus, the free energy of corona $F_{corona}$ is given by

\begin{equation}
F_{corona}(m,p)=-\ln Z(m,p)  \label{Fcorona}
\end{equation}%
where $Z(m,p)$ is given by (\ref{star_partition_sat}) and we have
explicitly shown the parameter $p$ that will be relevant in the
following discussion. Implicit in eq. (\ref{Fcorona}) is the fact
that the size of the hydrophilic block is much larger than that of
the hydrophobic, so that the micelle can be regarded as
effectively a star.

This repulsive contribution of the corona, which tends to solubilize the
copolymers is balanced by the attraction of hydrophobic units in the core.
The core contribution $F_{core}(p)$ can be written as

\begin{equation}
F_{core}(p)=4\pi R_{c}^{2}(p)\gamma
\end{equation}
where $\gamma $ is the surface tension between the core and the solvent,
being the core of size $R_{c}$. Assuming dense packing of the monomers in
the core, the radius of the core composed by $p$ block copolymers can be
expressed as

\begin{equation}
R_{c}(p)=\left[ v\frac{3}{4\pi }pL_{c}\right] ^{\frac{1}{3}}
\end{equation}
where $v$ is the volume of a monomer, $L_{c}$ is the length of the
hydrophobic block. Finally, a contribution, $F_{conf}(p)$ should be added
due to the entropy reduction due to the aggregation of $p$ chains into a
micelle, yielding

\begin{equation}
F_{conf}(p)=(p-1)\ln \left[ \frac{p}{V_{agg}e}v\right]
\end{equation}%
See Appendix \ref{secappendmic} for the details. We consider that the core
is compact and thus $V_{agg}\sim R_{c}^{3}$; hence $V_{agg}\sim vpL_{c}$.

Hence, the free energy of a micelle of $p$ block copolymers and $m$ stickers
in the corona is the sum
\begin{equation}
F(m,p)=F_{corona}(m,p)+F_{core}(p)+F_{conf}(p)
\end{equation}%
where $p\geq 1$. The particular case $F(0,p)$ is the free energy of a clean
(with no colloids adhered) micelle of $p$ block copolymers, while $F(m,1)$
is the free energy of a free chain decorated with $m$ colloids. To establish
the equilibrium, we have to further introduce the free energy of isolated
colloids which read, $F_{colloid}=c_{f}\ln (\frac{c_{f}}{e}\Lambda_s^{3})$,
where $c_{f}$ stands for the concentration of free colloids in the bulk and $%
\Lambda_s^{3}$ is the de Broglie's length of the colloid. Notice that we
consider that the bulk solution of colloids is dilute and only the
translational entropy is relevant. Furthermore, the concentration of free
chains and micelles of any size are also sufficiently dilute as to neglect
interactions among them.

Let $c(m,p)$ be the number concentration of micelles with $p$ arms and $m$
colloids. Hence, the total free energy of the solution containing all kinds
of aggregates is

\begin{eqnarray}
\frac{F}{VkT} &=&\sum_{m=0}^{1/\alpha }\sum\limits_{p=1}^{\infty }\left(
c(m,p)\ln \frac{c(m,p)v}{e}+c(m,p)\left(F(m,p)-f_{ref}(T) \right)\right)
+c_{f}\ln \frac{c_{f}v_s}{e} -c_{f}f_s(T)  \notag \\
&&+\mu _{0}\left( c_{0}-\sum_{m=0}^{1/\alpha }\sum\limits_{p=1}^{\infty
}mc(m,p)-c_{f}\right) +\mu _{b}\left( c_{b}-\sum_{m=0}^{1/\alpha
}\sum_{p=1}^{\infty }pc(m,p)\right)  \label{micellization}
\end{eqnarray}%
where the last two terms fix the total amount of colloids and copolymer
chains in the system, respectively. $V$ is the volume of the system, $%
c_{0}\equiv N_{0}/V$ and $c_{b}\equiv N_{b}/V$, where $N_{0}$ and $N_{b}$
are, respectively, the total number of colloids and copolymers in the
system. Furthermore, $f_s(T)$ stands for $\ln v_s/\Lambda_s^3$. In turn, $%
f_{ref}(T) \equiv \ln v/\lambda^3$ (see appendix \ref{secappendmic}).
Minimization of this free energy with respect to $c(m,p)$ gives the
equilibrium distribution of the aggregates by their size $p$ and number of
adhered colloids, $m$. That is,

\begin{equation}
vc(m,p)=(vc(0,1))^{p}(v_{s}c_{f})^{m}\exp \left( -\left[
F(m,p)-f_{ref}-p(F(0,1)-f_{ref})-m(F(1,0)-f_{s})\right] \right)
\end{equation}%
where the Lagrange multipliers $\mu _{0}$ and $\mu _{b}$ have been expressed
through the concentrations of unimers of each species, $c_{f}$ for free
colloids and $c(0,1)$ for free and clean block copolymers, which can be both
considered as the control parameters.

To obtain the relationship between volume fractions of the species, we
assume that there is no volume of mixing in the system. Then, the
equilibrium distributions are given by
\begin{eqnarray}
\phi (m,p) &=&\phi ^{p}(0,1)\phi _{f}^{m}\frac{Npv+mv_{s}}{N^{p}v}\times
\notag \\
&&\exp \left[ -\left(
F(m,p)-f_{ref}-p(F(0,1)-f_{ref})-m(F(1,0)-f_{s})\right) \right]
\end{eqnarray}%
where $\phi (m,p)$ is the volume fraction of a micelle of $p$ arms with its
adhered $m$ stickers, while $\phi (0,1)$ and $\phi _{s}$ are, respectively,
the volume fractions of free polymer and free colloid.

To determine the effect of the presence of the adhered colloids on the
micellization properties, in the following we shall focus on the behavior of
the function,

\begin{eqnarray}
\Omega (m,p) &=&\ln \frac{c(0,1)}{c(m,p)}=  \notag \\
&=&-(p-1)\ln (c(0,1)v)-m\ln (c_{f}v_{s})+F^{\ast }(m,p)-pF^{\ast }(0,1)
\label{Omega}
\end{eqnarray}%
whose minima coincide with the maxima of $c(m,p)$ and, therefore, gives us
the average size of the micelle. Here $F^{\ast }(m,p)\ \equiv F(m,p)-f_{ref}$
for any $p\leq 1$ and we consider no energy associated with colloids.

\begin{figure}[th]
\begin{center}
\includegraphics[width=8.5cm]{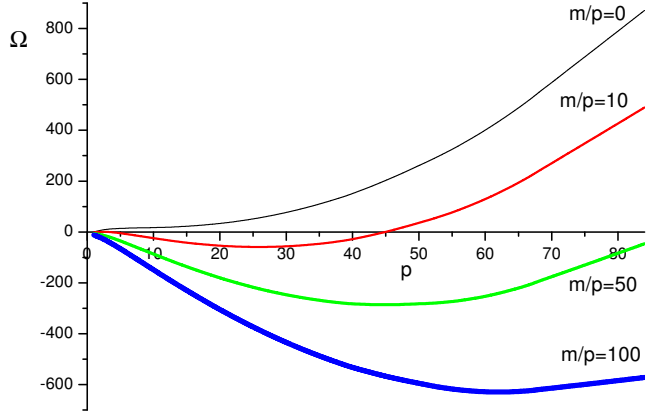}
\end{center}
\caption{{Variation of }$\Omega (m,p)$ as a function of $p$ for different
values of $m/p$. $c_{f}$ has been set to $e^{-10}$ and $c(0,1)$, to $e^{-30}$%
, $\protect\gamma =0.7$, $L_{c}=50$ and $L=1000$. Parameters $y\equiv e^{-%
\protect\varepsilon /kT}\Delta ^{\protect\sigma _{4}-d\protect\nu +1}=25$
and $\protect\alpha \equiv \Delta /L=10^{-4}$. Notice that for $c(0,1)$ is
below the CMC in the absence of colloids. The presence of colloids favors
the micellization as well as the formation of larger aggregates as larger is
$m/p$.}
\label{FigOmega}
\end{figure}

The convenience of this function is the possibility of naturally defining
the critical aggregation concentration (cac), i.e., when the concentration
of unimers and aggregates is of the same order, this potential is close to $%
0 $, while in the absence of the aggregation $\Omega (m,p)$ goes to
infinity. For convenience, we analyze $\Omega (m,p)$\ as a function of $p$
for fixed $m/p$, to see how the presence of the adhered colloids change the
properties of the "renormalized" block copolymer with $m$ colloids per
branch. In Figure \ref{FigOmega} we make this analysis for different values
of $m/p$. As we see the presence of adhered colloids induces the
micellization and the growth of the micelles as $m/p$ increased.

We  have focused our analysis on the regime where colloids are
rather scarce along the polymer chains and the polymer
concentration remains small in the associated micelles. Under
these conditions the obtained micelle coronas are in the  critical
excluded volume regime and the bare interaction between effective
monomers (sensitive to the presence of adsorbed colloids) is
unimportant. Association is then favored by the effective
shortening of the arms. However, at least close to the (small)
core of the micelle, there may exist a region where the polymer is
in the mean-field regime and therefore be sensitive to the bare
magnitude of the excluded volume interaction. The effect of
colloids on micellization then may be inverted there, that is,
that colloid adhesion does not favor micellization. This regime
deserves further consideration, although it lies beyond the scope
of this work. Finally, if the chains are heavily loaded in
colloids and lose flexibility the excluded volume interaction
drops and micellization is then favored by the colloids.

\section{Conclusions \label{secconclusion}}

A scaling theory is constructed aiming at the study of the conformations of
hydrophilic polymers interacting with small colloidal particles that can
reversibly stick onto a polymer backbone. Different geometries have been
considered, namely linear polymer chains, star polymers, as well as coronas
of block copolymer micelles. Unlike related analyses, we have consistently
taken into account the integration of the position of the colloid over the
chain backbone, due to the reversible adhesion of the former onto the
polymer.

We find that small colloids adhered to a given polymer induce formation of
independent loops of minimal size on each colloid, which can slide along the
chain backbone, thus increasing the entropy gain in comparison with static
loops of large size. This result is independent of the geometry of the
polymer as well as of the nature of the interactions. The sliding degree of
freedom of such small loops provides the system with an additional entropic
contribution proportional to the length of the polymeric branch. More
complex structures such as combined loops, ailed loops, or cactus-like
structures are not favorable because of the lack of such a sliding degree of
freedom. This contribution has been disregarded in previous analyses in
similar systems that rely on the scaling of networks of fixed topology. This
is one of the main results of this work.

We have applied these concepts to the equilibrium between colloids and star
polymers finding that the saturation of the polymer by the colloids is a
crossover process, rather than a phase transition, except for the case of a
star with infinitely long arms, where the crossover turns into higher order
phase transition in the sense of Ehrenfest, that is, with a finite
discontinuity in the third derivative of the free energy. Moreover, the
analysis of the micellization problem in the presence of colloids agrees
well with the observed experimental fact that the adhesion of colloids
favors both, the micellization process, as well as the growth of larger
micelles.

\acknowledgments{The authors acknowledges the financial support
from Spanish Ministry of education MICINN via project
CTQ2008-06469/PPQ. The authors are grateful to Dr. F. Siperstein
for inspiration of this work.}

\appendix

\section{Ailed loop\label{secailedloop}}

To illustrate this fact, we can for example calculate the partition function
of ailed loops created by the presence of two stickers in the chain, as
shown in Figure \ref{fig2}. Similar structures occur in knots formed by slip
link contacts \cite{Metzler2}.

Denoting by $a$ the length of the common section of the two ailed loops,
whose total sizes are $n_{1}$ and $n_{2}$ respectively, and $l$ is the size
of the shortest tail, we can write the partition function as $Z\sim
a^{u}n_{1}^{x_{1}}n_{2}^{x_{2}}l^{y}(N-l-n_{1}-n_{2}-a)^{z}$.

Without loss of generality we choose $n_{1}<n_{2}$, since the loops are
symmetrical. In analogy with the previous case it can be shown that the
configurations with the loops fixed at a free end, when the short tail
satisfies $l<n_{1}<n_{2}$ are subdominant. Then, we consider $n_{1}<n_{2}<l$%
. If the common section is small, $a<n_{1}<n_{2}<l$ we can fix the
exponents. If all segments are of the same order of magnitude and comparable
to the total size of the chain, one has

\begin{equation}
u+x_{1}+x_{2}+y+z=2\sigma _{1}+2\sigma _{4}-2d\nu
\end{equation}
If the size of the common segment decreases to one single monomer, $%
a\rightarrow 1$,

\begin{equation}
x_{1}+x_{2}+y+z=2\sigma _{1}+\sigma _{6}-2d\nu
\end{equation}
If one loop is negligibly small, $a<n_{1}\rightarrow 1$, then

\begin{equation}
x_{2}+y+z=2\sigma _{1}+\sigma _{4}-d\nu
\end{equation}
When both loops are vanishing, $a<n_{1}<n_{2}\rightarrow 1$,

\begin{equation}
y+z=2\sigma _{1}
\end{equation}
And, finally, small tail is vanishing,

\begin{equation}
z=2\sigma _{1}
\end{equation}
From these equations all exponents can be calculated and the total partition
function takes this form

\begin{eqnarray}
Z &\sim &\int_{0}^{N/2}dl\int_{0}^{l}dn_{2}n_{2}^{\sigma _{4}-d\nu
}\int_{0}^{n_{2}}dn_{1}n_{1}^{\sigma _{6}-\sigma _{4}-d\nu }\times  \notag \\
&&\int_{0}^{n_{1}}daa^{2\sigma _{4}-\sigma _{6}}(N-l-n_{1}-n_{2}-a)^{2\sigma
_{1}}
\end{eqnarray}
After integration, we find

\begin{equation}
Z\sim \Delta ^{2(\sigma _{4}-d\nu )+3}N^{2\sigma _{1}}N
\end{equation}
The same scaling is obtained if one considers the other two possibilities,
namely, $n_{1}<n_{2}<a<l$ and $n_{1}<a<n_{2}<l$ and we have omitted its
explicit calculation. As we have seen, the entropy favors the smaller size
the possible for the loops. For this particular case this is obtained if
both loops tend to shrink to its minimal possible size determined by the
cut-off $\Delta $. However since by construction loops are ailed, they are
forced to move together along the chain backbone, therefore its entropy is a
factor $1/N$ smaller than if they could independently move. This kind of
reasoning can be extended to more complex topologies with the same result.

If the sliding link is formed by small associating colloidal particles that
can attach and detach from the chain at any point, the formation of
individual loops moving separately along the chain is more favorable
compared to such combined loops with common segments. Using the same
arguments one can see that the partition function of individual loops (\ref%
{mloops}) dominates any other structures with the same number of stickers
(combined loops, cactus structures of loops growing on top of each other,
etc.).

\section{Calculation of the entropy of the micelles\label{secappendmic}}

Self-assembly of polymers into finite size micelles is accompanied by the
entropy changes due to the confinement of the polymers assembled into the
micelle, which moves as a single object. Here, we estimate the entropic
contribution due to this confinement, which involves the integration of the
kinetic degrees of freedom, as well as the translational degree of freedom
of the polymers as a whole. We recall that the conformation degrees of
freedom have been explicitly considered in section \ref{secmodel}. We are
implicitly considering micellization as a two-state situation in which
polymers are either free or aggregated into a micelle containing $p$
polymers.

First, we calculate the entropy change due to the association of the $p$
polymers into a micelle in a system containing $n$ of such micelles, to
later introduce the effect of the presence of $m$ stickers in each of them.
Therefore, the partition function of $n$ non-interacting micelles comprised
of $p$ undiscernable polymers, each of length $N$, is written as

\begin{eqnarray}
Z &=&\frac{1}{n!}\left[ \frac{1}{p!}\int \frac{d\overrightarrow{p}_{1}\ldots
d\overrightarrow{p}_{Np}}{(2\pi \hbar )^{3pN}}\prod\limits_{i}^{pN}\exp
\left( -\frac{\overrightarrow{p}_{i}^{2}}{2m_{p}kT}\right) \times \right.
\notag \\
&&\left. \int d\overrightarrow{r}_{1}\ldots d\overrightarrow{r}_{p}\exp
\left( -\frac{U}{kT}\right) \right] ^{n}  \label{Zmain}
\end{eqnarray}%
where $m_{p}$ is the mass of a monomer, $U$ is the interaction potential
between polymers forming the micelle (undefined). The integration is over
all positions of the monomers $r$, and their momenta $p$. The micelles are
distinguishable only by their composition. We recall that the integration
over the translational degrees of freedom is constrained to the polymers to
move together as an aggregate, thus the separation between any pair of
monomers does not have to exeed the overall size of the aggregate.

Equation (\ref{Zmain}) can be rewritten in terms of the partition function
of a single micelle of $p$ polymers, $Z_{p}$, according to $Z=\frac{1}{n!}%
\left[ Z_{p}\right] ^{n}$, where $Z_{p}$ is written as

\begin{equation}
Z_{p}=\frac{1}{p!}\int \frac{d\overrightarrow{p}_{1}\ldots d\overrightarrow{p%
}_{Np}}{(2\pi \hbar )^{3pN}}\prod\limits_{i}^{pN}\exp \left( -\frac{%
\overrightarrow{p}_{i}^{2}}{2m_{p}kT}\right) \int d\overrightarrow{r}%
_{1}\ldots d\overrightarrow{r}_{p}\exp \left( -\frac{U}{kT}\right)
\end{equation}

The momentum integral can be readily calculated to give a term inversely
proportional to $\Lambda _{p}^{3pN}$, where $\Lambda _{p}$ is the de
Broglie's length of the monomer (thus, independent of $N$) $\Lambda
_{p}^{2}=h^{2}/(2\pi m_{p}kT)$, and $h$, Planck's constant. In the spatial
integration we will separate three contributions, namely, the overall
translation of the aggregate, related to the center of mass motion, the
relative motion of the polymers inside the aggregate, together with the
displacement of the monomers of every polymer. The latter gives a
contribution proportional to $l_{p}^{3(N-1)}$ per chain, where $l_{p}$ is a
length of the order of the radius of gyration of the chain, which we will
consider a molecular parameter. This contribution has effectively accounted
for in the free energy term of the micelle that we have previously
calculated (cf. sec. \ref{secmodel}). Thus, due to the fact that in sec. \ref%
{secmodel} the entropy terms are given in a dimensionless form, the factor $%
l_{p}^{3(N-1)}$ simply provides the appropriate dimensionality of the
expression. As far as the second contribution is concerned, the integration
is limited to a distance between any pair of polymers of the order of the
size of the aggregate, due to our initial hypothesis of the two-state system
(free and confined), and the description that we are doing here of the
precisely confined state. Therefore, this second contribution yields a term
proportional to $V_{agg}^{p-1}$, where $V_{agg}$ stands for the volume of
the aggregate. The remaining integration over the position of the center of
mass of the aggregate itself yields one additional factor proportional to
the overal volume $V$ of the system. Therefore, the final result reads

\begin{equation}
Z_{p}\sim \frac{1}{p!\Lambda _{p}^{3pN}}VV_{agg}^{p-1}l_{p}^{3p(N-1)}
\end{equation}

Hence, we can write this entropic contribution to the free energy of the
aggregate as

\begin{equation}
\frac{F}{kT}=-\ln Z\sim n\ln \left[ \frac{n}{Ve}\lambda ^{3}\right]
+n(p-1)\ln \left[ \frac{p}{V_{agg}e}\lambda ^{3}\right] -n\ln \frac{p}{e}
\label{Fid}
\end{equation}%
where $\lambda \equiv \Lambda _{p}^{N}/l_{p}^{N-1}$ summarizes only
molecular parameters. The first term is for the contribution of a ideal
solution of micelles. The second and the third terms stand for the entropy
penalty derived from keeping together $p$ polymers into an aggregate. The
third term is subdominant in the limit $p\gg 1$. If $p=1$ this expression
should recover the free energy of a dilute solution of independent polymers.
Thus, the last term, in addition, should be regarded as an artifact of the
distinction we have made between polymers inside the aggregate and outside
the aggregate. By consistency, the last term should be dropped to match that
mentioned limit, provided that in the more interesting situation $p\gg 1$ is
subdominant. Notice that the calculations done in sec. \ref{secmodel} take
into account the entropy due to the conformational changes of a system with $%
p$ arms, while here we are calculating the entropy penalty of putting these $%
p$ arms together into an aggregate, understanding that the aggregation
process is a physical process. Therefore, $V_{agg}$ should be regarded as
the volume of the core of the micelle and not that of the whole aggregate,
including the corona.

Similar arguments have been used in Refs. \cite{SemenovMic,SemenovPVP} to
derive the entropy penalty for the formation of the micelle, obtaining an
expression similar to the second term in our eq. (\ref{Fid}). The main
difference comes from the presence in our expression of the $\lambda ^{3}$
term. Therefore, the entropic contribution of Ref. \citenum{SemenovPVP}
should be regarded as with respect to a reference state. Effectively, let us
introduce a volume $V_{0}$, where $V_{0}\equiv Nv$, $v$ being the volume of
the monomer, into the logarithmic term, and substract it afterwards. The
free energy can be written as

\begin{equation}
\frac{F}{kT}=n\ln \left[ \frac{n}{eV}V_{0}\right] +n(p-1)\ln \left[ \frac{p}{%
eV_{agg}}V_{0}\right] -np\frac{f_{ref}}{kT}  \label{EqSemenov}
\end{equation}%
where $f_{ref}$ is the free energy per polymer in a reference state, defined
as
\begin{equation}
\frac{f_{ref}}{kT}=\ln \frac{V_0}{\lambda^3}
\end{equation}
Eq. (\ref{EqSemenov}) is exactly the same expression as in Ref. %
\citenum{SemenovMic}.

In the case of the colloids, the situation is different, for the entropy of
the adhered colloid has already taken into account in the derivation of $%
Z(m,p)$. Notice that this expression accounts for the energetic contribution
$\exp(-\varepsilon m)$, the translational entropy due to the colloid motion
along the chain backbone, as well as the indiscernibility of the colloids.
Therefore, we are only facing the contribution due to the chain aggregation.

Finally, for convenience, we introduce the volume of the monomer $v$ instead
of $V_0$ into the expression (\ref{Fid}). Then, eq. \ref{Fid} can be
rewritten as

\begin{equation}
\frac{F}{kT}=-\ln Z\sim n\ln \left[ \frac{n}{Ve}v\right] +n(p-1)\ln \left[
\frac{p}{V_{agg}e}v\right]-npf_{ref}(T)  \label{Ffin2}
\end{equation}
where $f_{ref}(T)\equiv \ln v/\lambda^3$ which is only dependent on the
temperature and molecular parameters, and, in addition, is independent of
the size of the polymer. We use this function in the construction of the
free energy in eq. (\ref{micellization}) for the systems of micelles of
different size and number of colloids in equilibrium.

Therefore, from this expression we can identify that the contribution that
yields $F_{conf}$ is given by the second term in \ref{Ffin2}, that is
\begin{equation}
n F_{conf}/kT=n(p-1)\ln \left[ \frac{p}{V_{agg}e}v\right]
\end{equation}

As it has been mentioned, the other contributions are explicitly accounted
for in the proposed expression for the free energy, eq. (\ref{micellization}%
).

\pagebreak
%\bibliographystyle{apsrev4-1}
%\bibliography{sticky}

\begin{thebibliography}{10}%
\makeatletter
\providecommand \@ifxundefined [1]{%
 \ifx #1\undefined \expandafter \@firstoftwo
 \else \expandafter \@secondoftwo
\fi
}%
\providecommand \@ifnum [1]{%
 \ifnum #1\expandafter \@firstoftwo
 \else \expandafter \@secondoftwo
\fi
}%
\providecommand \enquote [1]{``#1''}%
\providecommand \bibnamefont  [1]{#1}%
\providecommand \bibfnamefont [1]{#1}%
\providecommand \citenamefont [1]{#1}%
\providecommand\href[0]{\@sanitize\@href}%
\providecommand\@href[1]{\endgroup\@@startlink{#1}\endgroup\@@href}%
\providecommand\@@href[1]{#1\@@endlink}%
\providecommand \@sanitize [0]{\begingroup\catcode`\&12\catcode`\#12\relax}%
\@ifxundefined \pdfoutput {\@firstoftwo}{%
 \@ifnum{\z@=\pdfoutput}{\@firstoftwo}{\@secondoftwo}%
}{%
 \providecommand\@@startlink[1]{\leavevmode}%
 \providecommand\@@endlink[0]{}%
}{%
 \providecommand\@@startlink[1]{%
  \leavevmode
  \pdfstartlink
   attr{/Border[0 0 1 ]/H/I/C[0 1 1]}%
   user{/Subtype/Link/A<</Type/Action/S/URI/URI(#1)>>}%
  \relax
 }%
 \providecommand\@@endlink[0]{\pdfendlink}%
}%
\providecommand \url  [0]{\begingroup\@sanitize \@url }%
\providecommand \@url [1]{\endgroup\@href {#1}{\urlprefix}}%
\providecommand \urlprefix [0]{URL }%
\providecommand \Eprint[0]{\href }%
\@ifxundefined \urlstyle {%
  \providecommand \doi [1]{doi:\discretionary{}{}{}#1}%
}{%
  \providecommand \doi [0]{doi:\discretionary{}{}{}\begingroup
  \urlstyle{rm}\Url }%
}%
\providecommand \doibase [0]{http://dx.doi.org/}%
\providecommand \Doi[1]{\href{\doibase#1}}%
\providecommand \bibAnnote [3]{%
  \BibitemShut{#1}%
  \begin{quotation}\noindent
    \textsc{Key:}\ #2\\\textsc{Annotation:}\ #3%
  \end{quotation}%
}%
\providecommand \bibAnnoteFile [2]{%
  \IfFileExists{#2}{\bibAnnote {#1} {#2} {\input{#2}}}{}%
}%
\providecommand \typeout [0]{\immediate \write \m@ne }%
\providecommand \selectlanguage [0]{\@gobble}%
\providecommand \bibinfo [0]{\@secondoftwo}%
\providecommand \bibfield [0]{\@secondoftwo}%
\providecommand \translation [1]{[#1]}%
\providecommand \BibitemOpen[0]{}%
\providecommand \bibitemStop [0]{}%
\providecommand \bibitemNoStop [0]{.\EOS\space}%
\providecommand \EOS [0]{\spacefactor3000\relax}%
\providecommand \BibitemShut [1]{\csname bibitem#1\endcsname}%
%</preamble>
\bibitem{Hamley}%
  \BibitemOpen
  \bibfield{author}{%
  \bibinfo {author} {\bibfnamefont{I.}~\bibnamefont{Hamley}},\ }%
  \emph{\bibinfo {title} {Block Copolymers in Solution : Fundamentals and
  Applications}}\ (\bibinfo {publisher} {John Wiley \& Sons},\ \bibinfo {year}
  {2005})%
  \bibAnnoteFile{NoStop}{Hamley}%
\bibitem{Reiss}%
  \BibitemOpen
  \bibfield{author}{%
  \bibinfo {author} {\bibfnamefont{G.}~\bibnamefont{Riess}}, \bibinfo {author}
  {\bibfnamefont{G.}~\bibnamefont{Hurtrez}},\ and\ \bibinfo {author}
  {\bibfnamefont{P.}~\bibnamefont{Bahadur}},\ }%
  \emph{\bibinfo {title} {Block Copolymers}},\ \bibinfo {edition} {2nd}\ ed.,\
  Vol.~\bibinfo {volume} {2}\ (\bibinfo {publisher} {Wiley},\ \bibinfo
  {address} {New York},\ \bibinfo {year} {1985})%
  \bibAnnoteFile{NoStop}{Reiss}%
\bibitem{Halperin}%
  \BibitemOpen
  \bibfield{author}{%
  \bibinfo {author} {\bibfnamefont{T.~P.~L.}\ \bibnamefont{A.~Halperin},
  \bibfnamefont{M.~Tirrell}},\ }%
  \bibfield{journal}{%
  \bibinfo {journal} {Adv. Polym. Sci.}\ }%
  \textbf{\bibinfo {volume} {100}},\ \bibinfo {pages} {31} (\bibinfo {year}
  {1992})%
  \bibAnnoteFile{NoStop}{Halperin}%
\bibitem{Napper}%
  \BibitemOpen
  \bibfield{author}{%
  \bibinfo {author} {\bibfnamefont{D.}~\bibnamefont{Napper}},\ }%
  \emph{\bibinfo {title} {Polymeric stabilization of colloidal dispersions}}\
  (\bibinfo {publisher} {Academic Press},\ \bibinfo {address} {London},\
  \bibinfo {year} {1985})%
  \bibAnnoteFile{NoStop}{Napper}%
\bibitem{Kataoka}%
  \BibitemOpen
  \bibfield{author}{%
  \bibinfo {author} {\bibfnamefont{K.}~\bibnamefont{Kataoka}}, \bibinfo
  {author} {\bibfnamefont{A.}~\bibnamefont{Harada}},\ and\ \bibinfo {author}
  {\bibfnamefont{Y.}~\bibnamefont{Nagasaki}},\ }%
  \bibfield{journal}{%
  \bibinfo {journal} {Advanced Drug Delivery Reviews}\ }%
  \textbf{\bibinfo {volume} {47}},\ \bibinfo {pages} {113} (\bibinfo {year}
  {2001})%
  \bibAnnoteFile{NoStop}{Kataoka}%
\bibitem{Alberts}%
  \BibitemOpen
  \bibfield{author}{%
  \bibinfo {author} {\bibfnamefont{B.}~\bibnamefont{Alberts}}, \bibinfo
  {author} {\bibfnamefont{K.}~\bibnamefont{Roberts}}, \bibinfo {author}
  {\bibfnamefont{D.}~\bibnamefont{Bray}}, \bibinfo {author}
  {\bibfnamefont{J.}~\bibnamefont{Lewis}}, \bibinfo {author}
  {\bibfnamefont{M.}~\bibnamefont{Raff}},\ and\ \bibinfo {author}
  {\bibfnamefont{J.~D.}\ \bibnamefont{Watson}},\ }%
  \emph{\bibinfo {title} {The molecular biology of the cell}}\ (\bibinfo
  {publisher} {Garland},\ \bibinfo {address} {New York},\ \bibinfo {year}
  {1994})%
  \bibAnnoteFile{NoStop}{Alberts}%
\bibitem{Diez}%
  \BibitemOpen
  \bibfield{author}{%
  \bibinfo {author} {\bibfnamefont{A.}~\bibnamefont{Johner}}, \bibinfo {author}
  {\bibfnamefont{J.-F.}\ \bibnamefont{Joanny}}, \bibinfo {author}
  {\bibfnamefont{S.~D.}\ \bibnamefont{Orrite}},\ and\ \bibinfo {author}
  {\bibfnamefont{J.~B.}\ \bibnamefont{Avalos}},\ }%
  \bibfield{journal}{%
  \bibinfo {journal} {Europhys. Lett.}\ }%
  \textbf{\bibinfo {volume} {56}},\ \bibinfo {pages} {549} (\bibinfo {year}
  {2001})%
  \bibAnnoteFile{NoStop}{Diez}%
\bibitem{Gelbart}%
  \BibitemOpen
  \bibfield{author}{%
  \bibinfo {author} {\bibfnamefont{N.}~\bibnamefont{Bagatella-Flores}},
  \bibinfo {author} {\bibfnamefont{H.}~\bibnamefont{Schiessel}},\ and\ \bibinfo
  {author} {\bibfnamefont{W.~M.}\ \bibnamefont{Gelbart}},\ }%
  \bibfield{journal}{%
  \bibinfo {journal} {J. Phys. Chem. B}\ }%
  \textbf{\bibinfo {volume} {109}},\ \bibinfo {pages} {21305} (\bibinfo {year}
  {2005})%
  \bibAnnoteFile{NoStop}{Gelbart}%
\bibitem{Lykos}%
  \BibitemOpen
  \bibfield{author}{%
  \bibinfo {author} {\bibfnamefont{C.~N.}\ \bibnamefont{Likos}}, \bibinfo
  {author} {\bibfnamefont{C.}~\bibnamefont{Mayer}}, \bibinfo {author}
  {\bibfnamefont{E.}~\bibnamefont{Stiakakis}},\ and\ \bibinfo {author}
  {\bibfnamefont{G.}~\bibnamefont{Petekidis}},\ }%
  \bibfield{journal}{%
  \bibinfo {journal} {J. Phys.: Condens. Matter}\ }%
  \textbf{\bibinfo {volume} {17}},\ \bibinfo {pages} {S3363} (\bibinfo {year}
  {2005})%
  \bibAnnoteFile{NoStop}{Lykos}%
\bibitem{Diamant}%
  \BibitemOpen
  \bibfield{author}{%
  \bibinfo {author} {\bibfnamefont{H.}~\bibnamefont{Diamant}}\ and\ \bibinfo
  {author} {\bibfnamefont{D.}~\bibnamefont{Andelman}},\ }%
  \bibfield{journal}{%
  \bibinfo {journal} {Macromolecules}\ }%
  \textbf{\bibinfo {volume} {33}},\ \bibinfo {pages} {8050} (\bibinfo {year}
  {2000})%
  \bibAnnoteFile{NoStop}{Diamant}%
\bibitem{dynmesop}%
  \BibitemOpen
  \bibfield{author}{%
  \bibinfo {author} {\bibfnamefont{K.}~\bibnamefont{Flodstrom}}, \bibinfo
  {author} {\bibfnamefont{H.}~\bibnamefont{Wennerstrom}},\ and\ \bibinfo
  {author} {\bibfnamefont{V.}~\bibnamefont{Alfredson}},\ }%
  \bibfield{journal}{%
  \bibinfo {journal} {Langmuir}\ }%
  \textbf{\bibinfo {volume} {20}},\ \bibinfo {pages} {680} (\bibinfo {year}
  {2004})%
  \bibAnnoteFile{NoStop}{dynmesop}%
\bibitem{Iler}%
  \BibitemOpen
  \bibfield{author}{%
  \bibinfo {author} {\bibfnamefont{R.~K.}\ \bibnamefont{Iler}},\ }%
  \emph{\bibinfo {title} {The Chemistry of Silica}}\ (\bibinfo {publisher}
  {Wiley-Interscience},\ \bibinfo {address} {New York},\ \bibinfo {year}
  {1979})%
  \bibAnnoteFile{NoStop}{Iler}%
\bibitem{SilPEO}%
  \BibitemOpen
  \bibfield{author}{%
  \bibinfo {author} {\bibfnamefont{R.}~\bibnamefont{Takahashi}}, \bibinfo
  {author} {\bibfnamefont{K.}~\bibnamefont{Nakanishi}},\ and\ \bibinfo {author}
  {\bibfnamefont{N.}~\bibnamefont{Soga}},\ }%
  \bibfield{journal}{%
  \bibinfo {journal} {J. Sol-Gel Sci. Tech.}\ }%
  \textbf{\bibinfo {volume} {17}},\ \bibinfo {pages} {7} (\bibinfo {year}
  {2000})%
  \bibAnnoteFile{NoStop}{SilPEO}%
\bibitem{Micropor}%
  \BibitemOpen
  \bibfield{author}{%
  \bibinfo {author} {\bibfnamefont{M.}~\bibnamefont{Imperor-Clerc}}, \bibinfo
  {author} {\bibfnamefont{P.}~\bibnamefont{Davidson}},\ and\ \bibinfo {author}
  {\bibfnamefont{A.}~\bibnamefont{Davidson}},\ }%
  \bibfield{journal}{%
  \bibinfo {journal} {J. Am. Chem. Soc.}\ }%
  \textbf{\bibinfo {volume} {122}},\ \bibinfo {pages} {11925} (\bibinfo {year}
  {2000})%
  \bibAnnoteFile{NoStop}{Micropor}%
\bibitem{HaradaAcc}%
  \BibitemOpen
  \bibfield{author}{%
  \bibinfo {author} {\bibfnamefont{A.}~\bibnamefont{Harada}},\ }%
  \bibfield{journal}{%
  \bibinfo {journal} {Acc. Chem. Res.}\ }%
  \textbf{\bibinfo {volume} {34}},\ \bibinfo {pages} {456} (\bibinfo {year}
  {2001})%
  \bibAnnoteFile{NoStop}{HaradaAcc}%
\bibitem{HaradaCoo}%
  \BibitemOpen
  \bibfield{author}{%
  \bibinfo {author} {\bibfnamefont{A.}~\bibnamefont{Harada}},\ }%
  \bibfield{journal}{%
  \bibinfo {journal} {Coord. Chem. Rev.}\ }%
  \textbf{\bibinfo {volume} {148}},\ \bibinfo {pages} {115} (\bibinfo {year}
  {1996})%
  \bibAnnoteFile{NoStop}{HaradaCoo}%
\bibitem{HaradaLi3}%
  \BibitemOpen
  \bibfield{author}{%
  \bibinfo {author} {\bibfnamefont{A.}~\bibnamefont{Harada}}, \bibinfo {author}
  {\bibfnamefont{J.}~\bibnamefont{Li}},\ and\ \bibinfo {author}
  {\bibfnamefont{M.}~\bibnamefont{Kamachi}},\ }%
  \bibfield{journal}{%
  \bibinfo {journal} {J. Am. Chem. Soc.}\ }%
  \textbf{\bibinfo {volume} {116}},\ \bibinfo {pages} {3192} (\bibinfo {year}
  {1994})%
  \bibAnnoteFile{NoStop}{HaradaLi3}%
\bibitem{BaulinLayers}%
  \BibitemOpen
  \bibfield{author}{%
  \bibinfo {author} {\bibfnamefont{V.~A.}\ \bibnamefont{Baulin}}, \bibinfo
  {author} {\bibfnamefont{A.}~\bibnamefont{Johner}},\ and\ \bibinfo {author}
  {\bibfnamefont{C.~M.}\ \bibnamefont{Marques}},\ }%
  \bibfield{journal}{%
  \bibinfo {journal} {Macromolecules}\ }%
  \textbf{\bibinfo {volume} {38}},\ \bibinfo {pages} {1434} (\bibinfo {year}
  {2005})%
  \bibAnnoteFile{NoStop}{BaulinLayers}%
\bibitem{BaulinMic}%
  \BibitemOpen
  \bibfield{author}{%
  \bibinfo {author} {\bibfnamefont{V.~A.}\ \bibnamefont{Baulin}}, \bibinfo
  {author} {\bibfnamefont{N.-K.}\ \bibnamefont{Lee}}, \bibinfo {author}
  {\bibfnamefont{A.}~\bibnamefont{Johner}},\ and\ \bibinfo {author}
  {\bibfnamefont{C.~M.}\ \bibnamefont{Marques}},\ }%
  \bibfield{journal}{%
  \bibinfo {journal} {Macromolecules}\ }%
  \textbf{\bibinfo {volume} {39}},\ \bibinfo {pages} {871} (\bibinfo {year}
  {2006})%
  \bibAnnoteFile{NoStop}{BaulinMic}%
\bibitem{Duplantier}%
  \BibitemOpen
  \bibfield{author}{%
  \bibinfo {author} {\bibfnamefont{B.}~\bibnamefont{Duplantier}},\ }%
  \bibfield{journal}{%
  \bibinfo {journal} {J. Stat. Phys.}\ }%
  \textbf{\bibinfo {volume} {54}},\ \bibinfo {pages} {581} (\bibinfo {year}
  {1989})%
  \bibAnnoteFile{NoStop}{Duplantier}%
\bibitem{Mukamel}%
  \BibitemOpen
  \bibfield{author}{%
  \bibinfo {author} {\bibfnamefont{Y.}~\bibnamefont{Kafri}}, \bibinfo {author}
  {\bibfnamefont{D.}~\bibnamefont{Mukamel}},\ and\ \bibinfo {author}
  {\bibfnamefont{L.}~\bibnamefont{Peliti}},\ }%
  \bibfield{journal}{%
  \bibinfo {journal} {Phys. Rev. Lett.}\ }%
  \textbf{\bibinfo {volume} {85}},\ \bibinfo {pages} {4988} (\bibinfo {year}
  {2000})%
  \bibAnnoteFile{NoStop}{Mukamel}%
\bibitem{Metzler1}%
  \BibitemOpen
  \bibfield{author}{%
  \bibinfo {author} {\bibfnamefont{R.}~\bibnamefont{Metzler}},\ }%
  \bibfield{journal}{%
  \bibinfo {journal} {New Journal of Physics}\ }%
  \textbf{\bibinfo {volume} {4}},\ \bibinfo {pages} {91.1} (\bibinfo {year}
  {2002})%
  \bibAnnoteFile{NoStop}{Metzler1}%
\bibitem{Metzler3}%
  \BibitemOpen
  \bibfield{author}{%
  \bibinfo {author} {\bibfnamefont{A.}~\bibnamefont{Hanke}}\ and\ \bibinfo
  {author} {\bibfnamefont{R.}~\bibnamefont{Metzler}},\ }%
  \bibfield{journal}{%
  \bibinfo {journal} {Biophys. J}\ }%
  \textbf{\bibinfo {volume} {85}},\ \bibinfo {pages} {167} (\bibinfo {year}
  {2003})%
  \bibAnnoteFile{NoStop}{Metzler3}%
\bibitem{Vilar}%
  \BibitemOpen
  \bibfield{author}{%
  \bibinfo {author} {\bibfnamefont{J.~M.~G.}\ \bibnamefont{Vilar}}\ and\
  \bibinfo {author} {\bibfnamefont{L.}~\bibnamefont{Saiz}},\ }%
  \bibfield{journal}{%
  \bibinfo {journal} {Phys. Rev. Lett.}\ }%
  \textbf{\bibinfo {volume} {96}},\ \bibinfo {pages} {238103} (\bibinfo {year}
  {2006})%
  \bibAnnoteFile{NoStop}{Vilar}%
\bibitem{Berg}%
  \BibitemOpen
  \bibfield{author}{%
  \bibinfo {author} {\bibfnamefont{P.~H.}\ \bibnamefont{von Hippel}}\ and\
  \bibinfo {author} {\bibfnamefont{O.~G.}\ \bibnamefont{Berg}},\ }%
  \bibfield{journal}{%
  \bibinfo {journal} {J. Biol. Chem.}\ }%
  \textbf{\bibinfo {volume} {264}},\ \bibinfo {pages} {675} (\bibinfo {year}
  {1989})%
  \bibAnnoteFile{NoStop}{Berg}%
\bibitem{Broek}%
  \BibitemOpen
  \bibfield{author}{%
  \bibinfo {author} {\bibfnamefont{B.}~\bibnamefont{van~den Broek}}, \bibinfo
  {author} {\bibfnamefont{M.~A.}\ \bibnamefont{Lomholt}}, \bibinfo {author}
  {\bibfnamefont{S.-M.~J.}\ \bibnamefont{Kalisch}}, \bibinfo {author}
  {\bibfnamefont{R.}~\bibnamefont{Metzler}},\ and\ \bibinfo {author}
  {\bibfnamefont{G.~J.~L.}\ \bibnamefont{Wuite}},\ }%
  \bibfield{journal}{%
  \bibinfo {journal} {Proc. Nat. Acad. Sci.}\ }%
  \textbf{\bibinfo {volume} {105}},\ \bibinfo {pages} {15738} (\bibinfo {year}
  {2008})%
  \bibAnnoteFile{NoStop}{Broek}%
\bibitem{Poland}%
  \BibitemOpen
  \bibfield{author}{%
  \bibinfo {author} {\bibfnamefont{D.}~\bibnamefont{Poland}}\ and\ \bibinfo
  {author} {\bibfnamefont{H.}~\bibnamefont{Scheraga}},\ }%
  \bibfield{journal}{%
  \bibinfo {journal} {J. Chem. Phys.}\ }%
  \textbf{\bibinfo {volume} {45}},\ \bibinfo {pages} {1456} (\bibinfo {year}
  {1966})%
  \bibAnnoteFile{NoStop}{Poland}%
\bibitem{Metzler5}%
  \BibitemOpen
  \bibfield{author}{%
  \bibinfo {author} {\bibfnamefont{R.}~\bibnamefont{Metzler}}, \bibinfo
  {author} {\bibfnamefont{A.}~\bibnamefont{Hanke}}, \bibinfo {author}
  {\bibfnamefont{P.~G.}\ \bibnamefont{Dommersnes}}, \bibinfo {author}
  {\bibfnamefont{Y.}~\bibnamefont{Kantor}}, ,\ and\ \bibinfo {author}
  {\bibfnamefont{M.}~\bibnamefont{Kardar}},\ }%
  \bibfield{journal}{%
  \bibinfo {journal} {Phys. Rev. Lett.}\ }%
  \textbf{\bibinfo {volume} {88}},\ \bibinfo {pages} {188101} (\bibinfo {year}
  {2002})%
  \bibAnnoteFile{NoStop}{Metzler5}%
\bibitem{Ercolini}%
  \BibitemOpen
  \bibfield{author}{%
  \bibinfo {author} {\bibfnamefont{E.}~\bibnamefont{Ercolini}}, \bibinfo
  {author} {\bibfnamefont{F.}~\bibnamefont{Valle}}, \bibinfo {author}
  {\bibfnamefont{J.}~\bibnamefont{Adamcik}}, \bibinfo {author}
  {\bibfnamefont{G.}~\bibnamefont{Witz}}, \bibinfo {author}
  {\bibfnamefont{R.}~\bibnamefont{Metzler}}, \bibinfo {author}
  {\bibfnamefont{P.~D.~L.}\ \bibnamefont{Rios}}, \bibinfo {author}
  {\bibfnamefont{J.}~\bibnamefont{Roca}},\ and\ \bibinfo {author}
  {\bibfnamefont{G.}~\bibnamefont{Dietler}},\ }%
  \bibfield{journal}{%
  \bibinfo {journal} {Phys. Rev. Lett.}\ }%
  \textbf{\bibinfo {volume} {98}},\ \bibinfo {pages} {058102} (\bibinfo {year}
  {2007})%
  \bibAnnoteFile{NoStop}{Ercolini}%
\bibitem{Duplantier1}%
  \BibitemOpen
  \bibfield{author}{%
  \bibinfo {author} {\bibfnamefont{B.}~\bibnamefont{Duplantier}},\ }%
  \bibfield{journal}{%
  \bibinfo {journal} {Phys. Rev. Lett.}\ }%
  \textbf{\bibinfo {volume} {57}},\ \bibinfo {pages} {941} (\bibinfo {year}
  {1986})%
  \bibAnnoteFile{NoStop}{Duplantier1}%
\bibitem{Grassberger}%
  \BibitemOpen
  \bibfield{author}{%
  \bibinfo {author} {\bibfnamefont{H.-P.}\ \bibnamefont{Hsu}}, \bibinfo
  {author} {\bibfnamefont{W.}~\bibnamefont{Nadler}},\ and\ \bibinfo {author}
  {\bibfnamefont{P.}~\bibnamefont{Grassberger}},\ }%
  \bibfield{journal}{%
  \bibinfo {journal} {Macromolecules}\ }%
  \textbf{\bibinfo {volume} {37}},\ \bibinfo {pages} {4658} (\bibinfo {year}
  {2004})%
  \bibAnnoteFile{NoStop}{Grassberger}%
\bibitem{Binder}%
  \BibitemOpen
  \bibfield{author}{%
  \bibinfo {author} {\bibfnamefont{A.}~\bibnamefont{Baumgartner}}\ and\
  \bibinfo {author} {\bibfnamefont{K.}~\bibnamefont{Binder}},\ }%
  \bibfield{journal}{%
  \bibinfo {journal} {J. Chem. Phys.}\ }%
  \textbf{\bibinfo {volume} {71}},\ \bibinfo {pages} {2541} (\bibinfo {year}
  {1979})%
  \bibAnnoteFile{NoStop}{Binder}%
\bibitem{Everaersts}%
  \BibitemOpen
  \bibfield{author}{%
  \bibinfo {author} {\bibfnamefont{R.}~\bibnamefont{Everaersts}}, \bibinfo
  {author} {\bibfnamefont{I.~S.}\ \bibnamefont{Grahamtt}},\ and\ \bibinfo
  {author} {\bibfnamefont{M.~J.}\ \bibnamefont{Zuckermann}},\ }%
  \bibfield{journal}{%
  \bibinfo {journal} {J. Phys. A Math. Gen.}\ }%
  \textbf{\bibinfo {volume} {28}},\ \bibinfo {pages} {1271} (\bibinfo {year}
  {1995})%
  \bibAnnoteFile{NoStop}{Everaersts}%
\bibitem{GrassbergerDNA}%
  \BibitemOpen
  \bibfield{author}{%
  \bibinfo {author} {\bibfnamefont{M.~S.}\ \bibnamefont{Causo}}, \bibinfo
  {author} {\bibfnamefont{B.}~\bibnamefont{Coluzzi}},\ and\ \bibinfo {author}
  {\bibfnamefont{P.}~\bibnamefont{Grassberger}},\ }%
  \bibfield{journal}{%
  \bibinfo {journal} {Phys. Rev. E}\ }%
  \textbf{\bibinfo {volume} {62}},\ \bibinfo {pages} {3958} (\bibinfo {year}
  {2000})%
  \bibAnnoteFile{NoStop}{GrassbergerDNA}%
\bibitem{Ben-Naim}%
  \BibitemOpen
  \bibfield{author}{%
  \bibinfo {author} {\bibfnamefont{E.}~\bibnamefont{Ben-Naim}}, \bibinfo
  {author} {\bibfnamefont{Z.~A.}\ \bibnamefont{Daya}}, \bibinfo {author}
  {\bibfnamefont{P.}~\bibnamefont{Vorobieff}},\ and\ \bibinfo {author}
  {\bibfnamefont{R.~E.}\ \bibnamefont{Ecke}},\ }%
  \bibfield{journal}{%
  \bibinfo {journal} {Phys. Rev. Lett.}\ }%
  \textbf{\bibinfo {volume} {86}},\ \bibinfo {pages} {1414} (\bibinfo {year}
  {2001})%
  \bibAnnoteFile{NoStop}{Ben-Naim}%
\bibitem{Metzler2}%
  \BibitemOpen
  \bibfield{author}{%
  \bibinfo {author} {\bibfnamefont{R.}~\bibnamefont{Metzler}}, \bibinfo
  {author} {\bibfnamefont{A.}~\bibnamefont{Hanke}}, \bibinfo {author}
  {\bibfnamefont{P.~G.}\ \bibnamefont{Dommersnes}}, \bibinfo {author}
  {\bibfnamefont{Y.}~\bibnamefont{Kantor}},\ and\ \bibinfo {author}
  {\bibfnamefont{M.}~\bibnamefont{Kardar}},\ }%
  \bibfield{journal}{%
  \bibinfo {journal} {Phys. Rev. E}\ }%
  \textbf{\bibinfo {volume} {65}},\ \bibinfo {pages} {061103} (\bibinfo {year}
  {2002})%
  \bibAnnoteFile{NoStop}{Metzler2}%
\bibitem{SemenovMic}%
  \BibitemOpen
  \bibfield{author}{%
  \bibinfo {author} {\bibfnamefont{I.~A.}\ \bibnamefont{Nyrkova}}\ and\
  \bibinfo {author} {\bibfnamefont{A.~N.}\ \bibnamefont{Semenov}},\ }%
  \bibfield{journal}{%
  \bibinfo {journal} {Macromol. Theory Simul.}\ }%
  \textbf{\bibinfo {volume} {14}},\ \bibinfo {pages} {569} (\bibinfo {year}
  {2005})%
  \bibAnnoteFile{NoStop}{SemenovMic}%
\bibitem{SemenovPVP}%
  \BibitemOpen
  \bibfield{author}{%
  \bibinfo {author} {\bibfnamefont{I.~A.}\ \bibnamefont{Nyrkova}}\ and\
  \bibinfo {author} {\bibfnamefont{A.~N.}\ \bibnamefont{Semenov}},\ }%
  \bibfield{journal}{%
  \bibinfo {journal} {Faraday Discuss.}\ }%
  \textbf{\bibinfo {volume} {128}},\ \bibinfo {pages} {113} (\bibinfo {year}
  {2005})%
  \bibAnnoteFile{NoStop}{SemenovPVP}%
\end{thebibliography}

%

\end{document}